\documentclass[useAMS]{mn2e}
\usepackage{times}
\usepackage{rotate}
\usepackage{graphics}
\usepackage{psfig}
\usepackage{lscape}
\input{epsf}
\newif\ifAMStwofonts
\AMStwofontstrue

\def\gsim{\mathrel{\hbox{\rlap{\hbox{\lower4pt\hbox{$\sim$}}}\hbox{$>$}}}}
\def\lsim{\mathrel{\hbox{\rlap{\hbox{\lower4pt\hbox{$\sim$}}}\hbox{$<$}}}}

\def\chisq{{$\chi^{2}$}}

\title[X-ray Survey of the Groth Strip]{A Deep Chandra survey of the Groth Strip. I. The X-ray data}
\author[K. Nandra et al.]{K. Nandra$^{1}$, E.S. Laird$^{1}$, K. Adelberger$^{2}$,  Jonathan P. Gardner$^{3}$,  R.F. Mushotzky$^{4}$, \newauthor
 J. Rhodes$^{5}$, C.C. Steidel$^{5}$, H.I. Teplitz$^{6}$, K.A. Arnaud$^{4}$  \\
$^1$Astrophysics Group, Imperial College London, Blackett Laboratory,
Prince Consort Road, London SW7 2AW, UK \\ 
$^2$Carnegie Observatories, 813 Santa Barbara Street, Pasadena 91101, CA, USA \\
$^3$Laboratory for Astronomy and Solar Physics, Code 681, NASA's Goddard Space Flight Center, Greenbelt, MD 20771, USA \\
$^4$Laboratory for High Energy Astrophysics, Code 660, NASA's Goddard Space Flight Center, Greenbelt, MD 20771, USA \\
$^5$California Institute of Technology, Pasadena, CA 91125, USA \\
$^6$Spitzer Science Center, California Institute of Technology, Pasadena, CA 91125, USA \\
}
\date{}

\begin{document}

\maketitle
\label{firstpage}

\begin{abstract}
We present the results of a 200 ks Chandra observation of part of the Groth Strip region, using the ACIS-I instrument. We present a relatively simple method for the detection of point-sources and calculation of limiting sensitivities, which we argue is at least as sensitive and more self-consistent than previous methods presented in the literature. 158 distinct X-ray sources are included in our point-source catalogue in the ACIS-I area. The number counts show a relative dearth of X-ray sources in this  region. For example at a flux limit of $10^{-15}$~erg cm$^{-2}$ s$^{-1}$ around 20 per cent more soft band sources are detected in the HDF-N and almost $50$~per cent more in the ELAIS-N1 field, which we have analysed by the same method for comparison.  We find, however, that these differences are consistent with Poisson variations at $<2 \sigma$ significance, and therefore there is no evidence for cosmic variance based on these number counts alone. We determine the average spectra of the objects and find a marked difference between the soft-band selected sources, which have $\Gamma=1.9$ typical of unobscured AGN, and the hard-band selected sources, which have $\Gamma=1.0$. Reassuringly, the sample as a whole has a mean spectrum of $\Gamma=1.4\pm 0.1$, the same as the X-ray background. Nonetheless, our results imply that the fraction of sources with significant obscuration is only $\sim 25$~per cent, much less than predicted by standard AGN population synthesis models. 
This is confirmed by direct spectral fitting, with only a handful of objects showing evidence for absorption. After accounting for absorption, all objects are consistent with mean intrinsic spectrum of  $\Gamma=1.76 \pm 0.08$, very similar to local Seyferts. The survey area is distinguished by having outstanding multi-waveband coverage. Comparison with these observations and detailed discussion of the X-ray source properties will be presented in future papers. 
 \end{abstract}


\section{INTRODUCTION}
\label{Sec:Introduction}

Deep X-ray surveys, particularly with Chandra, have had enormous recent successes in resolving the X-ray background and shedding light on the nature and evolution of accretion power in galaxies (e.g. Mushotzky et al. 2000; Giacconi et al. 2001; Barger et al. 2001, 2002; Cowie et al. 2002, 2003). It should be borne in mind, however, that the number of fields for which both deep Chandra data and comprehensive supporting mutliwavelength observations exists are relatively few. Aside from the deepest surveys in the HDF-N (Brandt et al. 2001; Alexander et al. 2003 hereafter A03) and CDF-S (Giacconi et al. 2001, 2002; Tozzi et al. 2001; Rosati et al. 2002), only two other fields have been presented with a Chandra exposure in excess of 150 ks (Stern et al. 2002; Wang et al. 2004).

A particularly well studied field in wavebands outside the X-ray is the Groth/Westphal Survey area (hereafter GWS). The original ``Groth Strip" survey with HST (Groth et al. 1995)  was complemented by a single deeper pointing, and further data have been obtained as part of the Canada-France redshift survey (CFRS; Lilly et al. 1995). Comprehensive imaging with HST/ACS has now also been approved. The CFRS data have themselves been superceded by the Canada-France Deep Field (CFDF) survey (McCracken et al. 2001) and together they provide multi-band imaging over much of the field.  It has also been covered by deep optical imaging, and followup spectropscopy, in a Lyman Break galaxy survey by Steidel et al. 2003. A further large optical spectroscopy effort is being undertaken as part of the DEEP and DEEP2 galaxy redshift surveys (Davis et al. 2003). SCUBA has observed a large area of the Groth Strip to deep limits for the Canada-UK Deep submm survey (Eales et al. 2000; Webb et al. 2003), and there is also coverage by ISO (Flores et al. 1999) and in the radio (Fomalont et al. 1991).  With a large number of ongoing or future programs for galaxy evolution in this field (e.g. DEEP2, Spitzer, CFHT, HST/ACS), and current interest in the connection between AGN and galaxy formation and evolution, deep X-ray data are clearly of great value. 

Some X-ray data have been obtained in this field with XMM (Waskett et al. 2003; Miyaji et al. 2004). Here we report Chandra observations which probe $\sim 5$ times deeper in flux, obtained in a 200 ks exposure with the ACIS-I CCD camera as prime instrument. This is currently the third deepest Chandra exposure, after the HDF-N and CDF-S. 

\section{OBSERVATIONS AND DATA REDUCTION}

\begin{table*}
\centering
\caption{Log of Chandra observations
Col.(1): Sequence number of the observation;
Col.(2): Date of beginning of observations;
Col.(3): Time of beginning of observation;
Col.(4): Nominal right ascension of satellite pointing;
Col.(5): nominal declination;
Col.(6): Exposure time of unfiltered, level 1 events file  (ks);
Col.(7): Permitted background range (ct s$^{-1}$ - see text);
Col.(8): Exposure time after GTI, background filtering etc. (ks). Total good exposure was
190.6 ks;
\label{tab:obslog}}
\begin{center}
\begin{tabular}{cccccccc}
\hline
Observation & Date & Time & RA & DEC & Raw & Background & Filtered \\
 ID & (UT) & (UT) & (J2000) & (J2000) & Exposure & Range & Exposure \\
(1) & (2) & (3) & (4) & (5) & (6)  & (7) & (8)\\
\hline
3305  & 2002-08-11 & 21:43:57 & 14:17:43.04 & 52:28:25.21 &  29.4 & 0.62--0.90 & 27.9 \\
4357  & 2002-08-12 & 22:32:00 & 14:17:43.04 & 53:28:25.20 &  86.1 & 0.60--0.87 & 80.5  \\
4365  & 2002-08-21 & 10:56:53 & 14:17:43.04 & 53:28:25.20 &  84.2 & 0.55--1.50 & 82.2 \\
\hline
\end{tabular}
\end{center}
\end{table*}

Chandra observed the GWS on three separate occasions between 2002-Aug-11 and 2002-Aug-22, using ACIS-I as the prime instrument. The S2 and S3 chips of the ACIS-S array were also operating during the observation but as these are far off-axis we do not consider the data further. The sequence number identifying the observations was 900144 and the three observation ID numbers (ObsIDs), along with some other basic observational information, are  given in Table~\ref{tab:obslog}.  The aim point of the observations were all very close to each other and close to the requested target position of $\alpha$ = 14:17:43.6, $\delta=$52:28:41.2. The latter is the position of the deep ``Westphal'' pointing of HST/WFPC2 and the center of the Lyman Break Galaxy survey field of Steidel et al. (2003).  Our data reduction was performed using the CXC Chandra Interactive Analysis of Observations (CIAO) data analysis software, version 3.0.1 and the Chandra calibration database (CALDB) version 2.23. 

We first used the CIAO aspect calculator to check for known aspect errors in the three observations. No such offset needed to be applied to the observations. The expectation is therefore that the raw Chandra astrometry should be good to 1 arcsec. CTI corrections and updated gain maps were applied to the unfiltered (level 1) event files. We applied the standard screening criteria to the observations, choosing event grades 0,2,3,4 \& 6 (the standard ``ASCA" grade set), applying the standard GTIs and removing flaring pixels and afterglow events. The filtered (level 2) event files were then energy selected to permit events only in the 0.5-8 keV range. Outside this energy range the background is greatly elevated relative to any source signal. 

During Chandra observations, it is known that there can be periods were the background is significantly higher compared to typical values. These background ``flares" are generally removed before analysis. To examine our observations for such effects, we initially performed a crude source detection on the 0.5-8 keV data for each ObsID separately using the {\tt celldetect} program, accepting all sources with signal-to-noise ratio (S/N) greater than 3. Excluding these detected sources, we generated background light curves for each observation.  Lightcurves of these source subtracted regions were analysed using the CIAO {\tt analyze\_ltcrv.sl} script, which flags the periods where the background was $\pm 3 \sigma$ from the mean  value. Each of the three observations required filtering according to this prescription - approximately 5.3 ks 
were filtered from observation 3305, 5.7 ks from observation 4357 and  25 ks from observation 4365. The final filtered exposure time after this procedure was 158.5 ks: thus around 20 per cent of the data was lost due to apparent background flares using the {\tt analyze\_ltcrv.sl} criteria.  
Examination of the light curves revealed that, rather than being primarily due to flaring, the bulk of the background rejection is due to a period of elevated, but relatively stable and well behaved background during observation 4365.  The average background for the source free regions was found to be $\sim 0.78$ ct s$^{-1}$ for the first 60 ks of that ObsID, but it rose to an average of $1.38$~ct s$^{-1}$ thereafter. Rather than adopt the {\tt analyze\_ltcrv.sl} criteria blindly, we have also defined subjective background criteria to exclude noisy periods for each observation. These are shown in Table~\ref{tab:obslog} and resulted in a final, total exposure time of 190.6ks. We verified post hoc that inclusion of the addtional $\sim 25$~ks of data with elevated background in ObsID 4365 does indeed increase the sensitivity. For example, using {\tt wavdetect} at a probability level of $10^{-7}$ results in 134 full-band sources using the full 190.6 ks (see below),
where only 127 sources with the more strictly screened 158.5 ks exposure.

Before co-adding, the relative astrometry was improved by registering the observations to the coordinate frame of the observation 4357. This was done using bright ($S/N > 4$) sources, within 6 arcmin of the aim point that were detected in each of the observations using the script {\tt align\_evt.pl}
(written by Tom Aldcroft). There were 7 such sources common to observations 4357 and 3305, and 15 common to 4357 and 4365. The data sets were then co-added using the  CIAO script {\tt merge\_all}. The resulting image was visually inspected to ensure  that merging had not caused smearing or double peaking of point sources. 

The 0.5-8 keV co-added level 2 event file was used to create images in  a number of energy bands. A specific issue in source detection is that of the upper energy bound. The HDF-N team has used an upper energy of 8 keV (e.g. Brandt et al. 2001), with the CDF-S team choosing to restrict their search to below 7 keV (Giacconi et al. 2002). In our data set we found the latter to be more sensitive during source detection.  More sources are detected in, e.g. the 0.5-7 keV image compared to the 0.5-8 keV image, and no extra source is detected in, e.g. the 4-8 keV image compared to the 4-7 keV image. We therefore restrict further analysis to the 0.5-7 keV (full band or FB), 0.5-2 keV (soft band or SB), 2-7 keV (hard band or HB) and 4-7 keV (ultra-hard band or UB) ranges. Source detection and other analysis proceeded with the images at the raw resolution of 0.492 arcsec per pixel. 

Effective exposure maps were created for each of these bands using the 
{\tt merge\_all} script, which takes into account the effects such as vignetting and gaps 
between chips. The maps were created at a single energy representative
of the mean energy of the photons in each band: 1 keV for the soft band, 
2.5 keV for the full, 4 keV for the hard and 5.5 keV for the ultra-hard.
These energies were determined post hoc by considering the average photon energy
of the detected sources in each band. The exposure maps cover the ACIS-I chips 0,1,2,3 
only, and are also unbinnned.

\begin{table*}
\centering
\caption{Number of detected sources.
Col.(1): Detection method;
Col.(2): Probability threshold;
Col.(3): Merged sources;
Col.(4); Full band sources;
Col.(5); Soft band sources;
Col.(6); Hard band sources;
Col.(7); Ultra-hard band sources;
Col.(8); Expected false sources per band for wavdetect this is $N_{pix} \times p_{\rm thresh}$. The expected number of false sources for our method is discussed in the text;
\label{tab:nsource}}
\begin{center}
\begin{tabular}{cccccccc}
\hline
Method & $p_{\rm thresh}$ & Merged & FB & SB & HB & UB & $N_{\rm false}$ \\
(1) & (2) & (3) & (4) & (5) & (6)  & (7) & (8) \\
This paper & $4\times 10^{-6}$  & 158 & 155 & 121 & 100 & 44 & 1.8 \\
wavdetect  & $1\times 10^{-7}$  & 142 & 134 & 114 & 86 & 41 & 1.8 \\
wavdetect & $1\times 10^{-6}$   & 170 & 151 & 132 & 101 & 48 & 17.7 \\
wavdetect & $1\times 10^{-5}$   & 251 & 193 & 167 & 130 & 68 & 177 \\
wavdetect & $1\times 10^{-4}$   & 894 & 379 & 414 & 317 & 274 & 1765 \\
\hline
\end{tabular}
\end{center}
\end{table*}

 \section{DATA ANALYSIS}

\begin{figure}
\scalebox{0.4}{{\includegraphics{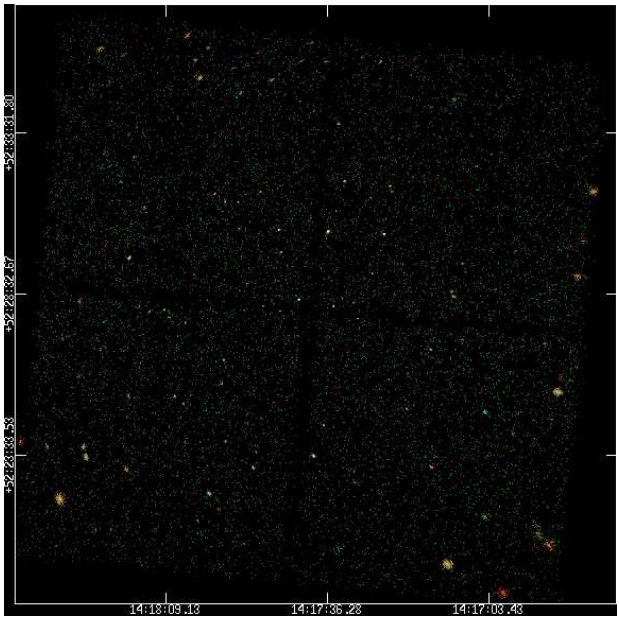}}}
\caption{Chandra 0.5-7 keV image of the GWS. The image is approximately
18 arcmin on each side
and has been smoothed with a Gaussian filter of 1 arcsec FWHM. The detected sources
are marked (green=detection in full band, red=soft band only). No source is detected
exclusively in the hard band, though many are hard band detections without significant
soft band counts. 
 \label{fig:xim}}
\end{figure}

Numerous different techniques have been applied to Chandra deep surveys to find sources, assess their significance, and calculate their fluxes and source counts. These tasks are complicated by the fact that the ACIS--I background is very low, meaning we have to deal with Poisson statistics. In addition, the point-source search algorithms often employed (e.g. {\tt wavdetect}, Freeman et al. 2002;  Sextractor Bertin \& Arnouts 1996) do not necessarily have well-defined sensitivities, making it difficult to define the flux limit as a function of area and/or assess the likely number of false sources in a catalogue. Generally, the tendency is to be conservative and try to minimize the latter. This has some advantages: in particular, it means that optical telescope time is not wasted following up ``spurious" sources. On the other hand, it runs the risk that many real sources may be missed. It is known that much information can be obtained about objects below these conservative thresholds, for example using fluctuation analysis (e.g. Miyaji \& Griffiths 2002) or stacking (e.g. Brandt et al. 2001; Nandra et al. 2002). In addition, different methods are often employed to identify likely significant sources, to determine their fluxes and errors and to determine the sensitivity limits (see Section~\ref{sec:comp}). This inconsistency has led us to develop our own point source detection procedure.

\subsection{Point source detection}

Our point source detection method is as follows: first, {\tt wavdetect} was run with a very low threshold probability of $10^{-4}$ on each of the four images, solely to identify positive fluctuations in the image which are candidate sources. A total of 1384 possible sources were identified in the four images, though many of these will have been detections in multiple bands. The numbers  for each individual band are shown in Table~\ref{tab:nsource}. We then extracted counts, again in each image separately, from circular apertures centered both on the {\tt wavdetect}-identified positions in that band and a centroided position, with an extraction radius equal to 70~per cent encircled energy of the Chandra Point Spread Function (PSF) at the appropriate mean energy of the image. 

To calculate the PSF radii we used the {\tt mkpsf} tool to create PSF images at the representative energy of each of the four bands, at the position of each candidate source, using the psf information in the CALDB PSF library {\tt acisi1998-11-052dpsf1N0002.fits}. We then extracted a radial profile of the PSF image by extracting counts from concentric circles centered on the source position out to a radius of 30 arcsec. Assuming all counts to be contained within the 30 arcsec radius, we then calculate by interpolation the radius at which 70~per cent of the counts are enclosed. This method can also be adapted to calculate the 90~per cent radius or other radii. 

We note that, particularly very far off axis in the hard energy band,  the assumption that all counts are contained within 30 arcsec may not strictly be valid, but any error introduced by this is likely to be very small. 

Background was determined locally from an annulus with inner radius equal to 1.5 times the 90~per cent PSF and outer radius 100 pixels greater than the inner radius. Detected sources from the {\tt wavdetect} run were excluded from the background region, as were zero-exposure pixels.
The background counts were then rescaled to the appropriate source region size, and by the ratio of the mean value of the exposure map in the source region to that in the background region. 
We can then calculate the Poisson probability that the source region would contain the number of counts it has based on the predicted background, apply a threshold, and determine what constitutes a detected source. This is performed for both the {\tt wavdetect} and centroid position of each candidate. If only one passes the threshold, we adopt that position. If both do, we adopt the position which gives the larger number of counts. 

We estimate the number of trials, and therefore the false source probability, by calculating
the number of the psf cells which fit into the total area of the image. This can be defined as: $$N_{\rm trial}=\sum_{i=1}^{N_{\rm pix}}\frac{a_{\rm pix}}{\pi r_{\rm PSF}^{2}}$$
where $a_{\rm pix}$ is the area of a pixel in arcsec and $r_{\rm PSF}$ is the 70 per cent radius
of the PSF in the same units, and the summation is over all the pixels in the image with non-zero exposure, $N_{\rm pix}$. For ultimate accuracy we would therefore have to calculate the PSF for $>4$M pixels.  The PSF calculation is time consuming, however, and this would be prohibitive. As we have already calculated the PSF for a large number of representative positions in the image -- for the candidate sources -- to extract their counts, we instead assume that the PSF radius for each pixel is equal to the closest of these positions (the ``nearest neighbour'' PSF). 

As the psf is energy dependent, $N_{\rm trial}$ depends on the energy band of the image. It ranges from $\sim 138,000$ for the SB, which has the smallest PSF, to $\sim 94,000$ for the UB, and the total of all four images was $4.6 \times 10^{5}$. The expected number of false sources is then $N_{\rm trial} \times p_{\rm thresh}$ where $p_{\rm thresh}$ is the threshold Poisson probability below which a sources is considered a detection. We adopt $p_{\rm thresh} = 4 \times 10^{-6}$, at which level we expect approximately 0.5 false sources in each of the four images, and 1.8 in total. The latter matches the expected number of false sources using {\tt wavdetect} with probability of $10^{-7}$, which is a typical value used to produce point source catalogues (e.g. A03; Wang et al. 2004). For a given $p_{\rm thresh}$, we expect fewer false sources with our method as we extract photons from cells with a fixed size and shape, unlike in {\tt wavdetect} where the extraction region is of arbitrary geometry (Freeman et al. 2002). 

\subsection{Flux determination}

The detection procedure described gives the total counts and background counts in the 70~per cent PSF region. To convert these to  for use in the logN-logS function, we background-subtracted the counts, divided by the average value of the exposure map in the source detection cell, and corrected for counts falling outside the cell. This gives count rates corrected for the majority of the instrumental effects, such as the chip gaps, mirror vignetting and the ACIS quantum efficiency. 

The CIAO exposure maps do not currently account for the degradation in the soft X-ray response of ACIS due to contamination (e.g. Marshall et al. 2004), the so-called ``ACISABS" correction. We determined this using the ACISABS model provided in the XSPEC spectral
fitting package.  
The correction factors for the counts were found to be 8.8 per cent, 13.9 per cent and 0.7 per cent for the FB, SB and HB respectively. The HB correction is therefore neglected, as is the negligible UB correction.
We calculated the fluxes correcting for the Galactic $N_{\rm H}$ of $1.3 \times 10^{20}$~cm$^{-2}$ and assuming a spectrum of $\Gamma=1.4$ for all the sources in all bands. This is the mean spectrum of the sources as deduced from analysis of the hardness ratios (see below). We converted the FB, HB and UB  counts to fluxes in standard bands: 0.5-10 keV, 2-10 keV and 5-10 keV respectively. The SB flux is quoted for the same band as the counts, 0.5-2 keV. 

For the source catalogue we want to merge sources detected in more than one band. The individual catalogues for the four bands were merged using a match radius of  $<2.5$ arcsec for sources within $6^{\prime}$ from the  average aim point. Sources with larger off-axis angle were matched with a radius of  $4$ arcsec to reflect the larger PSF. Our analysis indicates that the mismatch probability using these radii, over the whole field, is less than 1 per cent. We assign the position in the full band to the object if it is detected in that band, as it contains most counts and hence the smallest statistical uncertainty. If undetected in the FB, we adopted consecutively the SB, HB and UB positions, although in practice no object was detected only in the HB or UB. 

The 70 per cent PSF radius is appropriate for sensitive source detection, and we calculate the fluxes
from this for the logN-logS function because it maintains consistency between the source detection, the flux calculation and, as discussed below, the sensitivity calculation. It identifies positions in the image where there are significant sources of X-rays, and using a relatively small extraction radius optimises the sensitivity for source detection. Given the photon-starved nature of the Chandra images, however, it is preferable to use a larger radius for source photometry, to increase the number of counts and improve constraints on the flux. For the source catalogue, after band merging, we therefore also extracted the counts using a circular aperture of radius equal the point-spread function 90~per cent encircled energy fraction (FB, HB, UB), or 95~per cent (SB; following A03) and subtracted background counts from a surrounding annulus as described above. Errors on the counts were determine according to the prescription of Gehrels (1986). We used his equation (7) to calculate the effective ``1-sigma" upper bound on the counts and equation (14) for the corresponding lower bound. 

As the source detection procedure has ensured a very high probability that a real X-ray source exists at the identified position, there is no need to apply the same thresholding to the cross-band matches as has been applied to the source detection. Indeed, all that needs to be ensured is that there are sufficient counts to give a meaningful estimate of the flux in that band. Following this merging process, and again only for the source catalogue, we have re-extracted the counts in all the bands based on the merged position. If the significance in the band exceeds the probability equivalent to 3-sigma for a one tailed Gaussian distribution ($1.3 \times 10^{-3}$) we calculate the counts, fluxes and errors in that band. Otherwise we calculate the upper limit to the counts based on that same probability (i.e. the Poisson equivalent of a 3$\sigma$ upper limit), by determining the number of counts which would produce a probability less than that value given the background. 

\subsection{Sensitivity map}

The sensitivity map is needed  in order to calculate, for example, the logN-logS function for the field.  The aperture extraction procedure also allows us to determine the sensitivity map for these observations easily, and in a manner consistent with the source detection.

In order to calculate the sensitivity as a function of the survey area, and therefore the logN-logS, we  must calculate the counts detection threshold for an extraction radius centered at every point in the image. Again, we have adopted the ``nearest neighbour'' approach. For each pixel, we determine the closest candidate source.  We have already determined the PSF, and the expected background in the source cell for that position, with the latter simply needing to be rescaled to the mean exposure in the nominal source cell centered on each pixel. This rescaling will account, e.g. for the change in sensitivity due to the chip gaps and chip edges. With the given background, we then determine how many counts would constitute a detection at $<4 \times 10^{-6}$~Poisson probability, and convert that to the flux again using the mean exposure in the cell. We applied the ACISABS correction as above, and used a spectrum of $\Gamma=1.4$ to convert to flux, just as for the detected sources. 

%

\section{RESULTS}

\begin{figure}
\rotatebox{270}
{\scalebox{0.32}
{\includegraphics{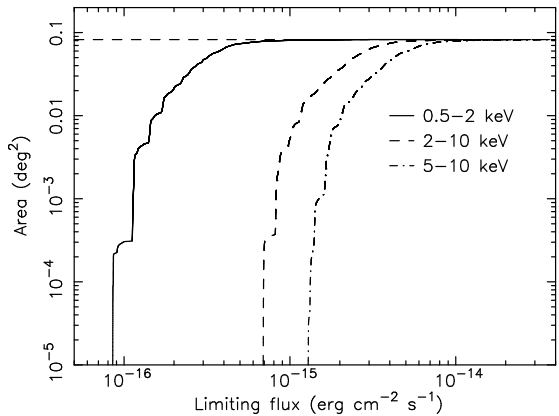}
}}
\caption{Limiting flux against survey area in the three bands for which we have calculated the logN-logS function, SB (0.5-2 keV; solid line), HB (2-10 keV; dashed line), UB (5-10 keB) dot-dash line. The total survey area is 0.082 deg$^{2}$. 
 \label{fig:slim}}
\end{figure}

\subsection{Point Source Catalogue}

The point source catalogue is shown in Table~\ref{tab:sample}. At our adopted detection threshold of $4 \times 10^{-6}$ a total of 158 band-merged sources are detected, with the numbers detected in the individual bands shown in Table~\ref{tab:nsource}. No object is detected solely in the HB or UB, although $\sim 25$~per cent of the sources remain undetected in the SB. These are candidate absorbed sources and we discuss their properties further below. In addition to the source counts, errors and detection probabilities in each band, the source catalogue also shows the fluxes in the standard bands, the off-axis angle and the hardness ratio, defined as: $$\frac{H-S}{H+S}$$ where H and S are the hard and soft band flux corrected to the on-axis value using  the instrument map.

\subsubsection{Comparison with {\tt wavdetect}}
\label{sec:comp}

The most common procedure for producing Chandra source catalogues it to use the the wavelet-based algorithm {\tt wavdetect} (Freeman et al. 2002). This algorithm is extremely powerful and effective method of finding sources in Chandra images. There are, however, some disadvantages to its use compared with our adopted method. For example, as we discuss below, it is not straightforward to construct a sensitivity map from a wavdetect run.  We first compare the efficiency of our source detection procedure to that of {\tt wavdetect}. 

We performed point source searches using the the CIAO {\tt wavdetect} agorithm on the four GWS images (full, soft, hard and ultrahard). Source detection was performed  for each of the four bands using the appropriate exposure map, using wavelet scales of 1, $\sqrt{2}$, 2, $2\sqrt{2}$, 
4, $4\sqrt{2}$, 8, $8\sqrt{2}$ and 16 pixels. The energy at which the PSF size was  
calculated in {\tt wavdetect} was set at 1 keV (SB), 2.5 keV (FB), 4 keV (HB) and 5.5 
keV (UB), just as in the creation of the exposure maps. Initially, the detection probability threshold was set at the typical value  $p_{\rm thresh} = 10^{-7}$. The {\tt wavdetect} documentation suggests that, as a rule of thumb, one would expect $p_{\rm thresh} \times N_{\rm pix}$ spurious sources in the detection run,  where $N_{\rm pix}$ is the total number of pixels in the image. The ACIS-I image of the GWS has a total of $4.41 \times 10^{6}$ pixels, so according to this prescription $\sim 0.44$ spurious sources are expected in each of the four images and 1.8 in all four. The total number of detected sources was 134 (FB), 114 (SB), 86 (HB) and 41 (UB). This compares to 155, 121, 100, 44 using our method (Table~\ref{tab:nsource}). Thus, our method produces more sources than {\tt wavdetect} at the probability threshold level which is generally used for ACIS-I images,  when we use a threshold expected to match {\tt wavdetect} in terms of false sources. 

Matching the catalogues shows that only two sources are detected in the {\tt wavdetect} run which are not identified in our procedure. Conversely 19 additional sources are detected with our method which are not found by {\tt wavdetect}. We have visually inspected all of these sources and are confident of their reality. Sometimes these new sources are at the edge of the chips (e.g. c2), or in the chip gaps (c68, c69). In one case (c52, c53) we find two sources very close together, while {\tt wavdetect} identifies only one as significant. Some caution needs to be applied in this case. 

We also investigated the effect of a lower $p_{\rm thresh}$ values by repeating the {\tt wavdetect} detection at  $p_{\rm thresh} = 10^{-6}, 10^{-5}$ and $10^{-4}$. The results are shown in Table~\ref{tab:nsource}. It can be seen that at a probability of $10^{-6}$ {\tt wavdetect} becomes
more sensitive than our method, but with an unacceptably high fraction ($>10$~per cent) of spurious sources. Interestingly, the expected number of spurious sources at the $10^{-4}$ probability level exceeds the total number of detections at that level (1384). As many of the latter are actually real sources, there is some evidence that the {\tt wavdetect} rule of thumb overestimates the number of false sources. If so,  {\tt wavdetect} could be run at probability levels lower then $10^{-7}$ without fear of excessive false detections, revealing the additional objects found in our analysis. The key point we make here, however, is that the additional sources we detect are very probably real, but would have been missed had we blindy used the typical {\tt wavdetect} threshold of $10^{-7}$

\subsubsection{Comparison fields}

We have also performed source detection for two comparison fields, the ISO ELAIS-N1 field, for which Chandra data have been presented previously by Manners et al. (2003; hereafter M03), and the 2Ms HDF-N (Alexander et al. 2003). Our standard screening procedure applied to ELAIS-N1 resulted in an exposure of 70ks. The HDF-N analaysis is described in more detail by Laird et al. (2004), and is more complicated than that described above as there are multiple pointings with very different aimpoints and roll angles. Our detection procedure resulted in a total of 145 sources for ELAIS-N1, and 536 for the HDF-N applying a threshold probability of $4\times 10^{-6}$ as for the GWS. Again these compare favourably to {\tt wavdetect}. M03 report 127 ACIS-I sources in ELAIS-N1 and A03 503 for HDF-N. The analysis of the comparison fields is discussed  further in the Appendix.

\subsection{X-ray number counts}
\label{sec:lognlogs}

We have calculated the detection sensitivity as a function of area for 3 of the bands (SB, HB, UB) according to the prescription described above. This is shown in Figure~\ref{fig:slim}. The limiting fluxes in each band, defined as the flux to which at least $1$~per cent of the survey area is sensitive, are $1.1 \times 10^{-16}$~erg cm$^{-2}$ s$^{-1}$ (SB), $8.2 \times 10^{-16}$~erg cm$^{-2}$ s$^{-1}$ (HB) and $1.4 \times 10^{-15}$~erg cm$^{-2}$ s$^{-1}$ (UB). The total survey area is 0.082 deg$^{-2}$. The limiting flux in the full band is $3.5 \times 10^{-15}$~erg cm$^{-2}$ s$^{-1}$, which can be compared to the limit of the XMM survey given by Waskett et al. (2004) of approximately $2 \times 10^{-15}$~erg cm$^{-2}$ s$^{-1}$, i.e. the deepest part of our Chandra survey probes fluxes about 5--6 times deeper than the XMM survey. Averaged over the whole field, we detect about 3 times as many sources per unit area than the XMM survey, but note that the XMM field is around 3 times large than that of ACIS-I, so the total number of sources is very similar. 

\begin{figure}
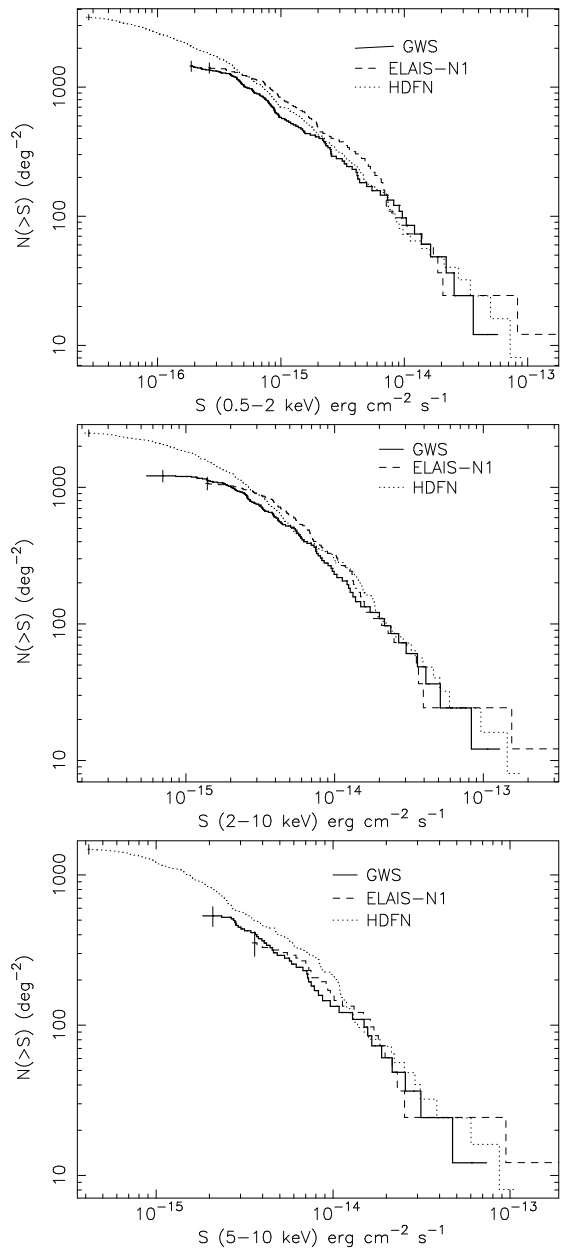

\rotatebox{270}
{\scalebox{0.32}
{\includegraphics{fig3a.ps}}
}
\rotatebox{270}
{\scalebox{0.32}
{\includegraphics{fig3b.ps}}
}
\rotatebox{270}
{\scalebox{0.32}
{\includegraphics{fig3c.ps}}
}
\caption{Cumulative logN-logS functions for the SB, HB and UB for the Groth strip and the two comparison fields (ELAIS-N1 and HDF-N). The faintest flux point for each field shows the Poisson
error bar. The source counts in all three bands are generally
lower in the GWS compared to the comparison fields, showing that it samples a relative void. However, even comparing to the field with the highest source density (ELAIS-N1), we find no significant evidence that the number counts differ beyond the expected Poisson variations (see text). 
The source counts show the effects of incompleteness at the faintest fluxes probed by
each survey. 
\label{fig:lnls}}
\end{figure}

The cumulative number count distributions (logN-logS) for the GWS  are shown in Figure~\ref{fig:lnls}. For comparison, we also show the distributions from the ELAIS-N1 deep survey (M03) and the HDF-N (Brandt et al. 2001; A03). The source density in the GWS is generally lower than in the other fields, which may indicate that we are sampling a relative void in the large scale structure. ``Cosmic variance'' of this kind has been suggested by previous investigations (e.g. Yang et al. 2003). For example, M03, comparing the ELAIS-N1 (analysed here) and ELAIS-N2 fields report that the former has $\sim 30$~per cent higher number counts, and attribute this to large scale clustering. We can investigate this with our data also. For example, comparing the soft X-ray logN-logS we can see that at a representative  flux level of $10^{-15}$, above the level where incompleteness and Eddington bias should seriously affect the results,  the ELAIS-N1 counts are $\sim 50$~per cent higher than those in the GWS. When we consider the expected Poissonian errors, however, we find the source densities of $827 \pm 100$ and $570 \pm 84$~deg$^{-2}$ are consistent at the 2$\sigma$ level. The differences in number counts in the hard band are less significant than this. 

\subsection{Spectral properties}

\subsubsection{Hardness ratios}

The hardness ratios of all objects in our sample are given in Table~\ref{tab:sample}, and our relatively lax criterion for cross-band detection means this quantity is defined for the great majority of sources. This allows us to calculate the mean hardness ratio for entire sample, and also for objects detected in the various sub-bands. As discussed by Nandra et al. (2003), when attempting to characterize the mean spectrum of an X-ray survey sample it is preferable to adopt the unweighted average of the hardness ratio. This is because using the weighted average, or count stacking can mean the result is dominated by a few bright sources. This is particularly important when it is suspected that the spectral properties might depend on flux, as has been suggested for Chandra samples (e.g. Giacconi et al. 2001). We find the following unweighted mean hardness ratio for sources detected in the various bands:  $-0.19 \pm 0.05$ (FB), $-0.42 \pm 0.03$ (SB), $-0.01\pm 0.05$ (HB), $-0.06\pm 0.07$ (UB). The hardness values in Table~\ref{tab:sample} have been calculated based on the counts corrected to on-axis values. They have not been corrected for the ACISABS effect, as we apply this when calculating the fluxes. We have considered this correction, however, when converting the hardness values to effective spectral indices, which we have performed using PIMMS. They are $\Gamma=1.4\pm0.1$ (FB - note that the same value is found when considering the whole sample of 158 sources), $\Gamma=1.85\pm 0.05$ (SB), $\Gamma=1.0\pm 0.1$ (HB) and $\Gamma=1.1\pm0.1$ (UB). There is clearly a highly significant difference in the mean spectrum of sources selected in the soft and hard bands  ($\sim 7.5\sigma$). 

\subsubsection{Spectral fitting}

\begin{figure}
\rotatebox{270}
{\scalebox{0.34}
{\includegraphics{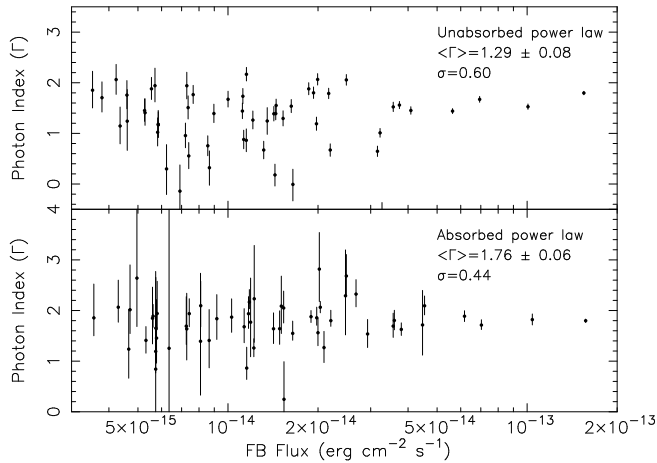}
}}
\caption{Power law photon index versus full band flux for the 56 sources with $>80$ total counts (FB) in a region defined by the 95 per cent PSF. The upper panel shows the results with Galactic $N_{\rm H}$ only. A wide dispersion of photon indices is observed, the data being completely inconsistent with a constant (reduced $\chi^{2}=12.8$). The mean index (unweighted) is consistent with that derived from the hardness ratio analysis and shows that these sources are representative of the full sample, and of the objects which make up the X-ray background. When additional absorption is included (lower panel), the dispersion reduces substantially, with the objects now being consistent with a single index of $\Gamma\sim 1.8$, similar to local Seyferts (Nandra \& Pounds 1994). Only a relatively small proportion of the objects require this additional absorption, however. Fluxes in the lower panel are corrected for intrinsic absorption. A single point with best-fit $\Gamma\sim7$ and very large error bars has been omitted for clarity. 
 \label{fig:spec}}
\end{figure}

In addition to hardness ratio analysis we have also performed direct spectral fitting on our sources. Spectra were extracted using the CIAO {\tt psextract} tool for the detected sources, using the 95 per cent PSF region for the full band, and the background using an annular region as described for the source detection. We performed spectral fitting only on sources with $>80$ source+background counts, to give at least 4 bins with a minimum of 20 counts per bin, permitting a $\chi^{2}$ fitting procedure. A total of 56 sources satisfied this criterion. We performed the fits in XSPEC, starting with a simple power law with Galactic absorption ($N_{\rm H}=1.27 \times 10^{20}$~cm$^{-2}$; Dickey \& Lockman 1990).  This proved a formally satisfactory fit in the great majority of cases, with only 9 sources being inconsistent with this model at $>99$~per cent confidence. The photon index ($\Gamma$) is plotted against flux in Fig.~\ref{fig:spec}. 

The objects display a very wide range of photon indices, from $\Gamma=-0.5$ to $2.2$. They are completely inconsistent with a single universal index, a hypothesis rejected at $>99.99$ per cent confidence. The unweighted average index, of $\Gamma=1.29\pm 0.08$ is very much consistent with that determined from the hardness ratio analysis so while the spectral subsample tends to be brighter it is representative of the sample as a whole. The weighted average index is $\Gamma=1.51\pm0.02$, slightly softer than the XRB but still consistent with the mean spectrum from the hardness ratio analysis. 

Nine of the objects show a spectrum inconsistent at $>99$~per cent confidence with a single power law absorbed by the Galactic $N_{\rm H}$ based on their \chisq. The most likely reason for this, and the hardness of the spectra, is absorption. We have tested this by introducing a neutral absorber into the model. As we do not have redshifts for the majority of the objects at present, we fix the absorber at z=0. This means that the derived column densities from the fits will not be accurate and we do not quote them. The fits should, however, give a reasonably accurate test of whether or not absorption is present in a given object, and of the effects of absorption on the power law index. This is because the photoelectric cross-section in the soft X-rays is dominated by K-shell absorption by moderate-z elements, which has an approximate $E^{-3}$ form for a wide range of atomic number. 

When absorption is allowed in the fit, a significant improvement (at $>99$ per cent confidence based on the F-test) is found in 8 objects. A more marginal improvement seen in some others is nonetheless likely to indicate the presence of absorption - for example c17 shows an acceptable fit when absorption is included where it was previously unacceptable. Indeed no source exhibits a spectrum inconsistent with this absorbed power law model at $>99$~per cent confidence. More complex models are not therefore required, nor would they be particularly meaningful without the redshifts. In the absorbed model the objects are now consistent with a universal index, of $\Gamma=1.76\pm 0.08$. 

Absorption clearly has a very substantial effect on the mean spectral properties, but as in the hardness ratio analysis the number of objects with significant absorption appears small. Only $\sim 15$~per cent of the spectral subsample (8/56) shows clear evidence for $N_{\rm H}$ above the Galactic value. Some of this may be due to poor spectral quality - the weakest source in which absorption is detected has $\sim 160$ counts, and we may simply be insensitive to absorption in objects fainter than this. Even if we restrict the analysis to objects with more counts than this limit, however, we still find only 25 per cent of objects (8/32) with significant absorption.

\section{DISCUSSION}

We have presented a source catalogue, number counts and spectral properties for the 200 ks deep survey of the GWS performed by Chandra. There has been a large variety of methods applied in the literature to identify sources, assign significances, determine fluxes and errors, and determining limiting sensitivities. Finding none of these entirely satisfactory, we have developed our own procedure based on pre-identification of candidate sources with {\tt wavdetect}, and a simple aperture extraction using the 70~per cent PSF and a local background to determine significances. This procedure is statistically well-defined, and it finds more sources than {\tt wavdetect} -- arguably the most common source detection procedure used in the literature. A total of 158 independent sources are detected in our band-merged catalogue, with the great majority ($>75$~per cent) being detected in the soft band and no source being detected exclusively in the hard or ultrahard bands. 

Observations have suggested significant differences in the number counts from field to field, which have been attributed to ``cosmic variance" (e.g. Cowie et al. 2002; Yang et al. 2003), but one study (Kim et al. 2004) has reported no such effect. Clearly it is important to investigate this issue further with deep survey data and we have done so with the GWS. While the logN-logS function shows lower counts than our comparison fields (by $\sim 50$~per cent compared to ELAIS-N1), we find the statistical significance of the difference in number counts to be weak ($<2\sigma$). This is not entirely surprising in that at the fluxes where the logN-logS is generally compared, the typical number of detected Chandra sources per Chandra field is very small (e.g. ~30 at $10^{-14}$erg cm$^{-2}$ s$^{-1}$. Large differences are therefore expected purely from Poisson variations. Greater source numbers - and hence lower Poisson noise - are present at fainter fluxes, but here the number counts appear to be consistent. As discussed by Yang et al., the main evidence for ``cosmic variance'' from the number counts comes from only 2 fields (including the CDF-S), which show low source numbers and apparently sample a relative void and only in the hard band.  It is clear also from the presence of ``spikes'' in the  redshift distribution of X-ray selected sources (e.g. Barger et al. 2002; Gilli et al. 2003) in deep surveys, that there must be significant field-to-field variations in the X-ray source population. We simply point out here that even large variations ($\sim 50$~per cent) in the number counts do not necessarily mean there is significant cosmic variance, and await further observations to confirm or deny the presence of these X-ray voids. 

Perhaps the most interesting result from our current analysis comes from the spectra. The whole sample (essentially the FB sources) shows a mean spectrum of $\Gamma=1.3-1.5$ from both the hardness ratio and direct spectral analysis.  This is consistent with that of the X-ray background, which is encouraging as our survey resolves the majority of that background. On the other hand we find a very marked difference between the mean spectra of soft and hard X-ray selected sources. The former have $\Gamma=1.9$. This mean spectrum is also in agreement with the average spectrum of (brighter) ROSAT-selected AGN (Georgantopoulous et al. 1997; Blair et al. 2000).  It is also typical of the intrinsic spectrum of local AGN (Nandra \& Pounds 1994). The implication is that these soft X-ray sources suffer essentially no obscuration. On the other hand the hard X-ray (2-7 keV) selected source show $\Gamma=1.0$, significantly flatter than the X-ray background. Once again this shows good consistency with brighter hard X-ray selected samples, e.g. the 5-10 keV selected SHEEP sample (Nandra et al. 2003). The surprise comes when we consider the fact that over 75~per cent of the whole sample are detected in the soft band, and hence relatively unobscured with mean $\Gamma=1.9$. Only a very small (25 per cent) population of heavily obscured objects detected in the hard band are therefore required to flatten the X-ray background to its observed level.

Direct spectral fitting of our Chandra data confirms this result, which is also found in XMM samples which probe slightly brighter fluxes (Piconcelli et al. 2003; Georgantopoulos et al. 2004). Our survey probes deeper into the populations producing the X-ray background and shows clearly and directly that the majority of objects producing that background are relatively unobscured in the X-ray. This is completely at odds with standard population synthesis models for the X--ray background (e.g. Comastri et al. 1995; Gilli et al. 1999, 2001) which predict that obscured sources should dominate. The basis of these synthesis models is the type I/II unification scheme which has proved so successful explaining the properties of local AGN (Antonucci \& Miller 1985).  Given the lack of correspondence between X-ray and optical measures of obscuration which is emerging in X-ray selected objects (e.g. Pappa et al. 2001, Comastri et al. 2002; Nandra et al. 2004), the basis of these synthesis models is also in doubt and they need to be examined completely afresh (see also Ueda et al. 2003). 

The spectral fitting shows that, after accounting for the effects of absorption in the few objects which require it, the objects in our sample are all consistent with a single power law slope, which has a mean of $\Gamma = 1.76 \pm 0.08$. This can be compared, for example, to the intrinsic spectrum of local Seyferts observed by Ginga, $\Gamma=1.95 \pm 0.05$ or ASCA, $\Gamma=1.91\pm 0.07$ (Nandra \& Pounds 1994; Nandra et al. 1997). These  include both the effects of (sometimes ionized) absorption and X-ray "Compton reflection". While the former may be crudely accounted for in our fits, the latter is not. If Compton reflection is present in the Chandra spectra - which the spectral quality does not allow us to identify - then the intrinsic spectrum of the objects is likely to be steeper by
$\Delta\Gamma \sim 0.1$.  Arguably the cleanest comparison is with the ASCA Seyferts without accounting for reflection, which show $\Gamma=1.79 \pm 0.07$, completely consistent with our derived Chandra mean. The existence of this ``canonical'' X-ray spectrum (e.g Turner \& Pounds 1989) argues for a common emission process for the X-rays in the entire AGN population. 

A final intriguing note is that the mean spectrum of the hard-band selected sources is consistent with that expected from one dominated by Compton-reflection (e.g. Matt et al.1996, 2000). Thus a significant population of ultra-obscured AGNs is not ruled out,  and indeed is supported by our analysis. Such objects are the likely origin of the peak in the X-ray background (Ueda et al. 2003), which occurs at $\sim 30$~keV (Marshall et al. 1984), and may make a significant contribution to the total discrete energy budget of the universe (Fabian \& Iwasawa 1999). The 75 per cent of our sample which are detected in the soft band, with typical $\Gamma=1.9$, are likely to make a negligible contribution to the XRB peak unless their spectrum changes radically as a function of energy. It is therefore possible that -- despite the extraordinary success of the Chandra deep surveys -- the majority of the energy density in the X-ray background  remains unresolved. The final mystery of the origin of the XRB therefore awaits deep surveys with a future, sensitive hard X-ray telescope. 

\section*{Acknowledgements}

This work has been supported by a Chandra grant awarded through SAO. We thank the Chandra X-ray Center for assistance with some analysis and software issues. ESL thanks PPARC for support in the form of a Research Studentship. 


\bsp

\label{lastpage}

\begin{landscape}
\begin{table}
\centering
\caption{Chandra GWS X-ray catalogue.
Col.(1): Source catalogue number;
Col.(2): Chandra object name;
Col.(3): Full band (0.5-7 keV) counts;
Col.(4): Soft band (0.5-2 keV) counts;
Col.(5): Hard band (2-7 keV) counts;
Col.(6): Ultra-hard band (4-7 keV) counts;
Col.(7): 0.5-10 keV flux (all fluxes $10^{-15}$ erg cm$^{-2}$ s$^{-1}$);
Col.(8): 0.5-2 keV flux;
Col.(9): 2-10 keV flux;
Col.(10): 5-10 keV flux;
Col.(11): $p_{\rm min}$ is the lowest false detection probability found for 
the four bands. Probabilities lower than $10^{-8}$ are not considered. 
Col.(12); Off-axis angle in arcmin. 
Col.(13); Hardness ratio = (H-S)/(H+S) where H and S are the 2-7 keV and 0.5- 2 keV counts, corrected to on-axis values. 
Col.(14); Flags fshu=source detected at
$<4 \times 10^{-6}$ probability in this band, where the bands are
f=full, s=soft, h=hard, u=ultrahard. The first band quoted is the one with the lowest
probability (i.e. that quoted in column 11). 
\label{tab:sample}}
\begin{center}
\begin{tabular}{@{}ccrrrrrrrrcccc@{}}
\hline
Cat & CXO GWS & FB & SB & HB & UB & 0.5-10.0 & 0.5-2.0 & 2-10 & 5-10 & 
  $p_{\rm min}$  & OAA & HR & Flags \\
No. & (2000) & cts & cts & cts & cts & flux & flux & flux & flux & & ($\prime$) & &  \\
(1) & (2) & (3) & (4) & (5) & (6) & (7) & (8) & (9) & (10) & (11) & (12) & (13) & (14)\\
\hline
c1    & J141642.1+523142 & $ 424.5^{+  22.4}_{ -21.4}     $ & $ 310.1^{+  19.0}_{ -18.0}     $ & $ 117.8^{+  12.8}_{ -11.8}     $ & $  36.9^{+   8.1}_{  -7.0}     $ & $ 41.00^{+  2.00}_{ -1.90}     $ & $ 12.00^{+  0.70}_{ -0.66}     $ & $ 25.00^{+  2.40}_{ -2.20}     $ & $ 17.00^{+  2.80}_{ -2.40}     $ &$10^{ -8.0}          $ &   9.90 &  -0.4 & fshu \\
c2    & J141643.5+522902 & $  26.8^{+   8.1}_{  -7.0}     $ & $  18.1^{+   6.3}_{  -5.2}     $ & $  10.2^{+   6.2}_{  -5.1}     $ & $< 12.13                       $ & $  2.90^{+  0.47}_{ -0.41}     $ & $  0.78^{+  0.18}_{ -0.15}     $ & $  2.50^{+  0.59}_{ -0.49}     $ & $<  6.00                       $ &$10^{ -6.5}          $ &   9.10 &  -0.3 & f    \\
c3    & J141644.0+523010 & $  66.0^{+  11.1}_{ -10.0}     $ & $  38.8^{+   8.4}_{  -7.3}     $ & $  27.6^{+   8.2}_{  -7.1}     $ & $< 13.96                       $ & $  4.80^{+  0.53}_{ -0.48}     $ & $  1.10^{+  0.17}_{ -0.15}     $ & $  4.50^{+  0.72}_{ -0.63}     $ & $<  4.60                       $ &$10^{ -8.0}          $ &   9.20 &  -0.2 & fsh  \\
c4    & J141645.3+522905 & $ 210.0^{+  16.6}_{ -15.6}     $ & $ 148.2^{+  13.8}_{ -12.7}     $ & $  64.0^{+  10.4}_{  -9.3}     $ & $< 13.97                       $ & $ 15.00^{+  1.00}_{ -0.94}     $ & $  4.10^{+  0.34}_{ -0.32}     $ & $ 10.00^{+  1.20}_{ -1.10}     $ & $<  4.40                       $ &$10^{ -8.0}          $ &   8.80 &  -0.4 & fsh  \\
c5    & J141646.9+523000 & $  43.1^{+   9.6}_{  -8.5}     $ & $  33.8^{+   7.9}_{  -6.8}     $ & $< 15.97                       $ & $< 13.25                       $ & $  2.90^{+  0.39}_{ -0.34}     $ & $  0.90^{+  0.15}_{ -0.13}     $ & $<  2.30                       $ & $<  4.00                       $ &$10^{ -8.0}          $ &   8.70 &  -1.0 & fs   \\
c6    & J141648.8+522558 & $  76.0^{+  11.2}_{ -10.2}     $ & $  54.3^{+   9.3}_{  -8.2}     $ & $  23.1^{+   7.5}_{  -6.4}     $ & $< 11.46                       $ & $  7.10^{+  0.77}_{ -0.70}     $ & $  2.00^{+  0.27}_{ -0.24}     $ & $  4.90^{+  0.89}_{ -0.76}     $ & $<  4.90                       $ &$10^{ -8.0}          $ &   8.60 &  -0.4 & fs   \\
c7    & J141649.4+522530 & $ 896.1^{+  31.4}_{ -30.4}     $ & $ 636.5^{+  26.5}_{ -25.5}     $ & $ 278.2^{+  18.2}_{ -17.2}     $ & $  75.3^{+  10.3}_{  -9.2}     $ & $ 87.00^{+  2.90}_{ -2.90}     $ & $ 24.00^{+  0.98}_{ -0.94}     $ & $ 61.00^{+  3.70}_{ -3.50}     $ & $ 35.00^{+  4.20}_{ -3.80}     $ &$10^{ -8.0}          $ &   8.70 &  -0.4 & fshu \\
c8    & J141651.2+522047 & $ 355.8^{+  21.2}_{ -20.2}     $ & $ 264.0^{+  18.0}_{ -17.0}     $ & $ 105.4^{+  12.7}_{ -11.7}     $ & $  27.2^{+   7.8}_{  -6.7}     $ & $ 30.00^{+  1.50}_{ -1.50}     $ & $  8.60^{+  0.54}_{ -0.51}     $ & $ 20.00^{+  1.80}_{ -1.70}     $ & $ 11.00^{+  2.00}_{ -1.70}     $ &$10^{ -8.0}          $ &  11.00 &  -0.4 & fsh  \\
c9    & J141652.0+522700 & $  57.4^{+   9.8}_{  -8.7}     $ & $  34.9^{+   7.6}_{  -6.5}     $ & $  25.4^{+   7.2}_{  -6.1}     $ & $< 10.93                       $ & $  4.00^{+  0.52}_{ -0.46}     $ & $  0.96^{+  0.17}_{ -0.15}     $ & $  4.00^{+  0.77}_{ -0.65}     $ & $<  3.50                       $ &$10^{ -8.0}          $ &   7.90 &  -0.1 & fsh  \\
c10   & J141653.4+522105 & $ 200.5^{+  16.5}_{ -15.5}     $ & $ 119.3^{+  12.8}_{ -11.8}     $ & $  82.8^{+  11.4}_{ -10.4}     $ & $  17.7^{+   6.7}_{  -5.6}     $ & $ 17.00^{+  1.20}_{ -1.10}     $ & $  4.00^{+  0.37}_{ -0.34}     $ & $ 16.00^{+  1.70}_{ -1.50}     $ & $  7.50^{+  1.60}_{ -1.30}     $ &$10^{ -8.0}          $ &  10.50 &  -0.2 & fsh  \\
c11   & J141653.8+522123 & $  53.8^{+  10.7}_{  -9.7}     $ & $< 15.49                       $ & $  44.2^{+   9.4}_{  -8.3}     $ & $  18.8^{+   6.9}_{  -5.8}     $ & $  4.10^{+  0.47}_{ -0.43}     $ & $<  0.47                       $ & $  7.70^{+  1.00}_{ -0.92}     $ & $  7.20^{+  1.50}_{ -1.20}     $ &$10^{ -8.0}          $ &  10.30 &   1.0 & fhu  \\
c12   & J141658.5+522412 & $  40.8^{+   8.5}_{  -7.5}     $ & $  20.3^{+   6.3}_{  -5.2}     $ & $  21.5^{+   6.7}_{  -5.6}     $ & $<  9.53                       $ & $  3.10^{+  0.47}_{ -0.41}     $ & $  0.60^{+  0.14}_{ -0.11}     $ & $  3.60^{+  0.76}_{ -0.64}     $ & $<  3.20                       $ &$10^{ -8.0}          $ &   8.00 &   0.1 & fsh  \\
c13   & J141659.1+522241 & $  19.3^{+   7.3}_{  -6.2}     $ & $   4.1^{+   4.7}_{  -3.5}     $ & $  14.6^{+   6.4}_{  -5.3}     $ & $< 11.17                       $ & $  1.40^{+  0.27}_{ -0.23}     $ & $  0.12^{+  0.04}_{ -0.03}     $ & $  2.50^{+  0.56}_{ -0.46}     $ & $<  3.80                       $ &$10^{ -5.8}          $ &   8.80 &   0.6 & f    \\
c14   & J141659.2+523436 & $  20.3^{+   7.5}_{  -6.4}     $ & $< 11.94                       $ & $  16.6^{+   6.6}_{  -5.5}     $ & $   9.0^{+   5.3}_{  -4.2}     $ & $  1.40^{+  0.25}_{ -0.21}     $ & $<  0.32                       $ & $  2.60^{+  0.55}_{ -0.46}     $ & $  2.90^{+  0.87}_{ -0.68}     $ &$10^{ -5.9}          $ &   9.10 &   1.0 & fh   \\
c15   & J141700.0+522304 & $  20.9^{+   7.2}_{  -6.1}     $ & $< 11.05                       $ & $  15.2^{+   6.3}_{  -5.2}     $ & $< 10.35                       $ & $  1.60^{+  0.30}_{ -0.25}     $ & $<  0.33                       $ & $  2.60^{+  0.60}_{ -0.49}     $ & $<  3.50                       $ &$10^{ -8.0}          $ &   8.50 &   1.0 & fh   \\
c16   & J141700.6+521918 & $ 365.6^{+  21.4}_{ -20.4}     $ & $ 309.1^{+  19.2}_{ -18.2}     $ & $  59.7^{+  10.7}_{  -9.7}     $ & $  21.3^{+   7.4}_{  -6.3}     $ & $ 29.00^{+  1.50}_{ -1.40}     $ & $  9.70^{+  0.56}_{ -0.53}     $ & $ 11.00^{+  1.20}_{ -1.10}     $ & $  8.40^{+  1.50}_{ -1.30}     $ &$10^{ -8.0}          $ &  11.20 &  -0.7 & fsh  \\
c17   & J141704.1+522140 & $  69.4^{+  10.6}_{  -9.6}     $ & $  21.9^{+   6.7}_{  -5.6}     $ & $  47.5^{+   9.0}_{  -7.9}     $ & $  17.4^{+   6.3}_{  -5.2}     $ & $  4.90^{+  0.56}_{ -0.51}     $ & $  0.60^{+  0.13}_{ -0.11}     $ & $  7.40^{+  1.10}_{ -0.93}     $ & $  5.80^{+  1.30}_{ -1.10}     $ &$10^{ -8.0}          $ &   9.00 &   0.4 & fshu \\
c18   & J141704.2+522453 & $ 107.2^{+  11.9}_{ -10.8}     $ & $  12.2^{+   5.2}_{  -4.1}     $ & $  98.5^{+  11.3}_{ -10.2}     $ & $  48.7^{+   8.3}_{  -7.2}     $ & $  7.20^{+  0.72}_{ -0.66}     $ & $  0.32^{+  0.10}_{ -0.08}     $ & $ 15.00^{+  1.60}_{ -1.40}     $ & $ 15.00^{+  2.30}_{ -2.00}     $ &$10^{ -8.0}          $ &   6.90 &   0.8 & fshu \\
c19   & J141705.7+523146 & $  16.9^{+   6.0}_{  -4.9}     $ & $<  7.50                       $ & $  13.3^{+   5.3}_{  -4.2}     $ & $<  6.97                       $ & $  1.30^{+  0.33}_{ -0.26}     $ & $<  0.23                       $ & $  2.30^{+  0.67}_{ -0.53}     $ & $<  2.30                       $ &$10^{ -8.0}          $ &   6.60 &   1.0 & fh   \\
c20   & J141705.7+523230 & $  35.4^{+   7.7}_{  -6.6}     $ & $  27.1^{+   6.6}_{  -5.5}     $ & $   9.0^{+   5.0}_{  -3.8}     $ & $<  7.92                       $ & $  2.60^{+  0.46}_{ -0.39}     $ & $  0.79^{+  0.17}_{ -0.14}     $ & $  1.50^{+  0.49}_{ -0.37}     $ & $<  2.50                       $ &$10^{ -8.0}          $ &   7.00 &  -0.5 & fs   \\
c21   & J141708.4+523225 & $  13.4^{+   5.7}_{  -4.5}     $ & $  10.8^{+   4.8}_{  -3.7}     $ & $<  8.75                       $ & $<  7.64                       $ & $  1.00^{+  0.27}_{ -0.22}     $ & $  0.31^{+  0.11}_{ -0.08}     $ & $<  1.40                       $ & $<  2.40                       $ &$10^{ -8.0}          $ &   6.60 &  -1.0 & sf   \\
c22   & J141708.6+522929 & $  39.0^{+   7.6}_{  -6.5}     $ & $  29.0^{+   6.6}_{  -5.5}     $ & $  11.2^{+   4.8}_{  -3.7}     $ & $<  6.08                       $ & $  2.80^{+  0.49}_{ -0.42}     $ & $  0.81^{+  0.17}_{ -0.14}     $ & $  1.80^{+  0.61}_{ -0.46}     $ & $<  1.80                       $ &$10^{ -8.0}          $ &   5.40 &  -0.4 & fs   \\
c23   & J141708.9+522708 & $  15.8^{+   5.7}_{  -4.5}     $ & $  11.7^{+   4.8}_{  -3.7}     $ & $   8.3^{+   4.6}_{  -3.4}     $ & $<  7.40                       $ & $  1.00^{+  0.28}_{ -0.22}     $ & $  0.30^{+  0.10}_{ -0.08}     $ & $  1.20^{+  0.46}_{ -0.34}     $ & $<  2.00                       $ &$10^{ -8.0}          $ &   5.30 &  -0.2 & f    \\
c24   & J141710.2+523433 & $  48.2^{+   9.1}_{  -8.0}     $ & $  21.0^{+   6.4}_{  -5.3}     $ & $  29.0^{+   7.3}_{  -6.2}     $ & $  10.6^{+   5.2}_{  -4.1}     $ & $  3.20^{+  0.46}_{ -0.40}     $ & $  0.55^{+  0.12}_{ -0.10}     $ & $  4.30^{+  0.81}_{ -0.69}     $ & $  3.30^{+  1.00}_{ -0.79}     $ &$10^{ -8.0}          $ &   7.90 &   0.2 & fshu \\
c25   & J141710.6+522828 & $ 119.4^{+  12.1}_{ -11.1}     $ & $  92.4^{+  10.7}_{  -9.7}     $ & $  29.5^{+   6.7}_{  -5.6}     $ & $<  6.32                       $ & $  8.50^{+  0.84}_{ -0.76}     $ & $  2.60^{+  0.30}_{ -0.27}     $ & $  4.60^{+  0.98}_{ -0.82}     $ & $<  1.90                       $ &$10^{ -8.0}          $ &   4.90 &  -0.5 & fsh  \\
c26   & J141711.0+522837 & $  40.7^{+   7.7}_{  -6.6}     $ & $  28.5^{+   6.6}_{  -5.4}     $ & $  14.7^{+   5.2}_{  -4.1}     $ & $   9.4^{+   4.4}_{  -3.3}     $ & $  3.00^{+  0.52}_{ -0.45}     $ & $  0.82^{+  0.18}_{ -0.15}     $ & $  2.40^{+  0.73}_{ -0.57}     $ & $  3.00^{+  1.20}_{ -0.90}     $ &$10^{ -8.0}          $ &   4.90 &  -0.3 & fshu \\
c27   & J141711.1+522541 & $  19.2^{+   6.0}_{  -4.9}     $ & $  19.8^{+   5.8}_{  -4.7}     $ & $<  7.63                       $ & $<  6.62                       $ & $  1.30^{+  0.33}_{ -0.27}     $ & $  0.54^{+  0.14}_{ -0.11}     $ & $<  1.10                       $ & $<  2.00                       $ &$10^{ -8.0}          $ &   5.60 &  -1.0 & fs   \\
c28   & J141711.6+523132 & $  17.5^{+   5.9}_{  -4.8}     $ & $  16.2^{+   5.4}_{  -4.3}     $ & $<  8.27                       $ & $<  6.47                       $ & $  1.10^{+  0.29}_{ -0.23}     $ & $  0.41^{+  0.12}_{ -0.09}     $ & $<  1.10                       $ & $<  1.70                       $ &$10^{ -8.0}          $ &   5.70 &  -1.0 & sf   \\
c29   & J141711.8+522011 & $ 950.9^{+  32.3}_{ -31.2}     $ & $ 720.2^{+  28.1}_{ -27.0}     $ & $ 268.9^{+  17.9}_{ -16.9}     $ & $  91.8^{+  11.1}_{ -10.1}     $ & $ 68.00^{+  2.20}_{ -2.20}     $ & $ 20.00^{+  0.77}_{ -0.74}     $ & $ 43.00^{+  2.70}_{ -2.50}     $ & $ 31.00^{+  3.40}_{ -3.10}     $ &$10^{ -8.0}          $ &   9.50 &  -0.4 & fshu \\
c30   & J141712.9+522207 & $  17.6^{+   6.6}_{  -5.4}     $ & $   5.3^{+   4.4}_{  -3.3}     $ & $  12.3^{+   5.7}_{  -4.5}     $ & $<  9.54                       $ & $  1.30^{+  0.28}_{ -0.24}     $ & $  0.15^{+  0.06}_{ -0.05}     $ & $  2.00^{+  0.54}_{ -0.44}     $ & $<  3.10                       $ &$10^{ -8.0}          $ &   7.80 &   0.4 & f    \\
c31   & J141714.3+522532 & $  14.8^{+   5.4}_{  -4.3}     $ & $  12.1^{+   4.8}_{  -3.7}     $ & $<  7.17                       $ & $<  6.09                       $ & $  0.96^{+  0.28}_{ -0.22}     $ & $  0.31^{+  0.11}_{ -0.08}     $ & $<  0.97                       $ & $<  1.60                       $ &$10^{ -8.0}          $ &   5.20 &  -1.0 & sf   \\
c32   & J141714.9+523420 & $  14.4^{+   6.2}_{  -5.1}     $ & $<  9.05                       $ & $  10.5^{+   5.3}_{  -4.2}     $ & $<  8.46                       $ & $  0.96^{+  0.23}_{ -0.19}     $ & $<  0.24                       $ & $  1.60^{+  0.46}_{ -0.36}     $ & $<  2.40                       $ &$10^{ -7.2}          $ &   7.30 &   1.0 & fh   \\
c33   & J141715.0+522312 & $ 149.0^{+  13.6}_{ -12.6}     $ & $ 112.8^{+  11.9}_{ -10.8}     $ & $  39.9^{+   7.8}_{  -6.8}     $ & $  10.2^{+   4.8}_{  -3.7}     $ & $ 10.00^{+  0.90}_{ -0.83}     $ & $  3.10^{+  0.32}_{ -0.29}     $ & $  6.20^{+  1.10}_{ -0.91}     $ & $  3.20^{+  1.10}_{ -0.84}     $ &$10^{ -8.0}          $ &   6.70 &  -0.5 & fshu \\
c34   & J141715.2+522649 & $  61.8^{+   9.1}_{  -8.0}     $ & $  19.7^{+   5.7}_{  -4.5}     $ & $  41.9^{+   7.7}_{  -6.6}     $ & $  11.5^{+   4.7}_{  -3.5}     $ & $  4.30^{+  0.60}_{ -0.53}     $ & $  0.54^{+  0.14}_{ -0.12}     $ & $  6.40^{+  1.10}_{ -0.96}     $ & $  3.50^{+  1.30}_{ -0.95}     $ &$10^{ -8.0}          $ &   4.50 &   0.4 & fshu \\
c35   & J141718.8+522743 & $   7.1^{+   4.0}_{  -2.8}     $ & $   6.6^{+   3.8}_{  -2.6}     $ & $<  4.33                       $ & $<  3.54                       $ & $  1.10^{+  0.53}_{ -0.37}     $ & $  0.39^{+  0.21}_{ -0.14}     $ & $<  1.40                       $ & $<  2.20                       $ &$10^{ -6.0}          $ &   3.70 &  -1.0 & f    \\
c36   & J141719.0+523051 & $  11.4^{+   4.8}_{  -3.7}     $ & $   7.8^{+   4.1}_{  -2.9}     $ & $<  6.11                       $ & $<  5.68                       $ & $  0.72^{+  0.25}_{ -0.19}     $ & $  0.19^{+  0.09}_{ -0.06}     $ & $<  0.80                       $ & $<  1.50                       $ &$10^{ -8.0}          $ &   4.40 &  -1.0 & fs   \\
c37   & J141719.3+522755 & $  14.7^{+   5.1}_{  -4.0}     $ & $   9.4^{+   4.3}_{  -3.1}     $ & $   5.0^{+   3.6}_{  -2.4}     $ & $<  4.37                       $ & $  1.50^{+  0.47}_{ -0.36}     $ & $  0.37^{+  0.16}_{ -0.12}     $ & $  1.10^{+  0.67}_{ -0.44}     $ & $<  1.80                       $ &$10^{ -8.0}          $ &   3.60 &  -0.3 & fs   \\
c38   & J141720.0+522500 & $  47.6^{+   8.2}_{  -7.1}     $ & $  27.5^{+   6.4}_{  -5.3}     $ & $  22.6^{+   6.1}_{  -5.0}     $ & $   4.3^{+   3.6}_{  -2.4}     $ & $  3.10^{+  0.49}_{ -0.43}     $ & $  0.70^{+  0.16}_{ -0.13}     $ & $  3.20^{+  0.78}_{ -0.64}     $ & $  1.20^{+  0.73}_{ -0.48}     $ &$10^{ -8.0}          $ &   4.90 &  -0.1 & fsh  \\

\hline
\end{tabular}
\end{center}
\end{table}
\end{landscape}

\setcounter{table}{2}
\begin{landscape}
\begin{table}
\centering
\caption{Chandra GWS X-ray catalogue (continued)}
\begin{center}
\begin{tabular}{@{}ccrrrrrrrrcccc@{}}
\hline
Cat & CXO GWS & FB & SB & HB & UB & 0.5-10.0 & 0.5-2.0 & 2-10 & 5-10 & 
  $p_{\rm min}$ & OAA& HR & Flags \\
No. & (2000) & cts & cts & cts & cts & flux & flux & flux & flux & & ($\prime$) & &  \\
(1) & (2) & (3) & (4) & (5) & (6) & (7) & (8) & (9) & (10) & (11) & (12) & (13) & (14) \\
\hline
c39   & J141720.4+522911 & $  21.4^{+   5.9}_{  -4.8}     $ & $  18.2^{+   5.4}_{  -4.3}     $ & $<  5.75                       $ & $<  5.04                       $ & $  1.30^{+  0.34}_{ -0.28}     $ & $  0.45^{+  0.13}_{ -0.10}     $ & $<  0.75                       $ & $<  1.30                       $ &$10^{ -8.0}          $ &   3.50 &  -1.0 & fs   \\
c40   & J141722.9+523143 & $  46.3^{+   8.1}_{  -7.0}     $ & $  35.7^{+   7.1}_{  -6.1}     $ & $  11.1^{+   4.7}_{  -3.5}     $ & $<  5.70                       $ & $  3.00^{+  0.50}_{ -0.43}     $ & $  0.93^{+  0.18}_{ -0.15}     $ & $  1.60^{+  0.58}_{ -0.44}     $ & $<  1.50                       $ &$10^{ -8.0}          $ &   4.50 &  -0.5 & fsh  \\
c41   & J141723.4+523153 & $ 216.3^{+  15.8}_{ -14.8}     $ & $ 176.7^{+  14.4}_{ -13.3}     $ & $  41.1^{+   7.6}_{  -6.5}     $ & $   9.7^{+   4.4}_{  -3.3}     $ & $ 15.00^{+  1.10}_{ -0.99}     $ & $  4.70^{+  0.38}_{ -0.35}     $ & $  6.10^{+  1.10}_{ -0.93}     $ & $  2.80^{+  1.10}_{ -0.84}     $ &$10^{ -8.0}          $ &   4.60 &  -0.6 & fshu \\
c42   & J141723.6+522555 & $  13.0^{+   5.0}_{  -3.8}     $ & $   4.1^{+   3.4}_{  -2.2}     $ & $   9.7^{+   4.4}_{  -3.3}     $ & $<  4.96                       $ & $  0.83^{+  0.27}_{ -0.21}     $ & $  0.10^{+  0.07}_{ -0.04}     $ & $  1.30^{+  0.54}_{ -0.40}     $ & $<  1.30                       $ &$10^{ -8.0}          $ &   3.90 &   0.4 & fh   \\
c43   & J141724.3+523229 & $  14.4^{+   5.3}_{  -4.2}     $ & $  12.2^{+   4.8}_{  -3.7}     $ & $<  7.38                       $ & $<  6.16                       $ & $  0.92^{+  0.27}_{ -0.21}     $ & $  0.31^{+  0.11}_{ -0.08}     $ & $<  0.98                       $ & $<  1.60                       $ &$10^{ -8.0}          $ &   5.00 &  -1.0 & fs   \\
c44   & J141724.6+523024 & $ 565.5^{+  24.8}_{ -23.8}     $ & $ 383.3^{+  20.6}_{ -19.6}     $ & $ 187.9^{+  14.8}_{ -13.7}     $ & $  56.1^{+   8.6}_{  -7.5}     $ & $ 36.00^{+  1.60}_{ -1.50}     $ & $  9.60^{+  0.52}_{ -0.49}     $ & $ 26.00^{+  2.00}_{ -1.90}     $ & $ 15.00^{+  2.30}_{ -2.00}     $ &$10^{ -8.0}          $ &   3.40 &  -0.3 & fshu \\
c45   & J141725.2+523512 & $  12.6^{+   6.0}_{  -4.9}     $ & $   9.2^{+   5.0}_{  -3.8}     $ & $< 10.57                       $ & $<  8.60                       $ & $  0.84^{+  0.21}_{ -0.17}     $ & $  0.24^{+  0.08}_{ -0.06}     $ & $<  1.50                       $ & $<  2.50                       $ &$10^{ -5.4}          $ &   7.30 &  -1.0 & f    \\
c46   & J141725.3+523544 & $  51.2^{+   9.2}_{  -8.1}     $ & $  27.2^{+   7.0}_{  -5.9}     $ & $  21.5^{+   6.6}_{  -5.5}     $ & $<  9.41                       $ & $  3.50^{+  0.48}_{ -0.42}     $ & $  0.72^{+  0.14}_{ -0.12}     $ & $  3.20^{+  0.69}_{ -0.58}     $ & $<  2.80                       $ &$10^{ -8.0}          $ &   7.80 &  -0.1 & fsh  \\
c47   & J141727.0+522911 & $  85.1^{+  10.3}_{  -9.2}     $ & $  62.5^{+   9.0}_{  -7.9}     $ & $  22.3^{+   5.9}_{  -4.8}     $ & $   8.4^{+   4.1}_{  -2.9}     $ & $  5.20^{+  0.63}_{ -0.56}     $ & $  1.50^{+  0.22}_{ -0.19}     $ & $  3.00^{+  0.76}_{ -0.62}     $ & $  2.20^{+  0.99}_{ -0.70}     $ &$10^{ -8.0}          $ &   2.60 &  -0.5 & fshu \\
c48   & J141727.3+523131 & $  11.9^{+   4.8}_{  -3.7}     $ & $   7.9^{+   4.1}_{  -2.9}     $ & $   6.5^{+   4.0}_{  -2.8}     $ & $<  4.93                       $ & $  0.75^{+  0.26}_{ -0.20}     $ & $  0.20^{+  0.09}_{ -0.06}     $ & $  0.90^{+  0.44}_{ -0.31}     $ & $<  1.20                       $ &$10^{ -8.0}          $ &   3.90 &  -0.1 & fs   \\
c49   & J141729.0+523553 & $  32.7^{+   7.8}_{  -6.8}     $ & $  21.8^{+   6.4}_{  -5.3}     $ & $  11.3^{+   5.6}_{  -4.4}     $ & $<  9.43                       $ & $  2.40^{+  0.41}_{ -0.35}     $ & $  0.63^{+  0.14}_{ -0.12}     $ & $  1.80^{+  0.51}_{ -0.41}     $ & $<  3.00                       $ &$10^{ -8.0}          $ &   7.80 &  -0.3 & fs   \\
c50   & J141729.9+522747 & $  87.5^{+  10.4}_{  -9.4}     $ & $  60.8^{+   8.9}_{  -7.8}     $ & $  28.6^{+   6.4}_{  -5.3}     $ & $   7.7^{+   4.0}_{  -2.8}     $ & $  9.20^{+  1.10}_{ -0.98}     $ & $  2.50^{+  0.37}_{ -0.32}     $ & $  6.60^{+  1.50}_{ -1.20}     $ & $  3.40^{+  1.70}_{ -1.20}     $ &$10^{ -8.0}          $ &   2.10 &  -0.3 & fshu \\
c51   & J141730.6+522242 & $  35.0^{+   7.5}_{  -6.4}     $ & $  25.1^{+   6.4}_{  -5.3}     $ & $  10.7^{+   5.0}_{  -3.8}     $ & $<  7.18                       $ & $  2.30^{+  0.43}_{ -0.36}     $ & $  0.66^{+  0.15}_{ -0.12}     $ & $  1.60^{+  0.52}_{ -0.40}     $ & $<  2.00                       $ &$10^{ -8.0}          $ &   6.00 &  -0.4 & fs   \\
c52   & J141730.6+522302 & $  20.9^{+   6.2}_{  -5.1}     $ & $  17.7^{+   5.6}_{  -4.4}     $ & $<  7.43                       $ & $<  6.63                       $ & $  1.40^{+  0.33}_{ -0.27}     $ & $  0.46^{+  0.13}_{ -0.10}     $ & $<  1.00                       $ & $<  1.80                       $ &$10^{ -8.0}          $ &   5.70 &  -1.0 & fs   \\
c53   & J141730.7+522305 & $  21.2^{+   6.2}_{  -5.1}     $ & $  16.6^{+   5.4}_{  -4.3}     $ & $   6.4^{+   4.3}_{  -3.1}     $ & $<  6.76                       $ & $  1.40^{+  0.33}_{ -0.27}     $ & $  0.43^{+  0.12}_{ -0.10}     $ & $  0.93^{+  0.40}_{ -0.29}     $ & $<  1.90                       $ &$10^{ -8.0}          $ &   5.60 &  -0.4 & fs   \\
c54   & J141730.8+522818 & $  16.2^{+   5.2}_{  -4.1}     $ & $<  3.68                       $ & $  12.4^{+   4.7}_{  -3.5}     $ & $   5.4^{+   3.6}_{  -2.4}     $ & $  0.99^{+  0.30}_{ -0.24}     $ & $<  0.09                       $ & $  1.70^{+  0.60}_{ -0.45}     $ & $  1.40^{+  0.83}_{ -0.55}     $ &$10^{ -8.0}          $ &   1.90 &   1.0 & fh   \\
c55   & J141732.6+523202 & $ 158.7^{+  13.7}_{ -12.7}     $ & $ 103.8^{+  11.3}_{ -10.2}     $ & $  59.5^{+   8.9}_{  -7.8}     $ & $  19.0^{+   5.6}_{  -4.4}     $ & $ 11.00^{+  0.93}_{ -0.86}     $ & $  2.80^{+  0.30}_{ -0.27}     $ & $  9.00^{+  1.30}_{ -1.10}     $ & $  5.60^{+  1.50}_{ -1.20}     $ &$10^{ -8.0}          $ &   4.00 &  -0.3 & fshu \\
c56   & J141733.6+522038 & $  41.5^{+   8.5}_{  -7.5}     $ & $<  9.55                       $ & $  35.1^{+   7.8}_{  -6.7}     $ & $  18.4^{+   6.0}_{  -4.9}     $ & $  2.90^{+  0.44}_{ -0.38}     $ & $<  0.26                       $ & $  5.40^{+  0.93}_{ -0.80}     $ & $  5.90^{+  1.50}_{ -1.20}     $ &$10^{ -8.0}          $ &   7.90 &   1.0 & fh   \\
c57   & J141733.8+523349 & $  64.6^{+   9.4}_{  -8.3}     $ & $  32.2^{+   7.0}_{  -5.9}     $ & $  32.2^{+   7.1}_{  -6.0}     $ & $  11.5^{+   4.8}_{  -3.7}     $ & $  4.20^{+  0.56}_{ -0.50}     $ & $  0.82^{+  0.16}_{ -0.14}     $ & $  4.60^{+  0.90}_{ -0.76}     $ & $  3.30^{+  1.10}_{ -0.86}     $ &$10^{ -8.0}          $ &   5.60 &   0.0 & fshu \\
c58   & J141734.0+522456 & $   8.1^{+   4.3}_{  -3.1}     $ & $<  5.15                       $ & $   7.5^{+   4.1}_{  -2.9}     $ & $   4.0^{+   3.4}_{  -2.2}     $ & $  0.51^{+  0.22}_{ -0.16}     $ & $<  0.13                       $ & $  1.00^{+  0.48}_{ -0.34}     $ & $  1.10^{+  0.74}_{ -0.47}     $ &$10^{ -6.1}          $ &   3.70 &   1.0 & fh   \\
c59   & J141734.4+523106 & $  33.7^{+   7.0}_{  -5.9}     $ & $  23.2^{+   6.0}_{  -4.9}     $ & $  10.1^{+   4.4}_{  -3.3}     $ & $<  4.29                       $ & $  2.20^{+  0.44}_{ -0.38}     $ & $  0.61^{+  0.15}_{ -0.12}     $ & $  1.50^{+  0.59}_{ -0.43}     $ & $<  1.10                       $ &$10^{ -8.0}          $ &   3.00 &  -0.4 & fsh  \\
c60   & J141734.8+522810 & $ 228.4^{+  16.2}_{ -15.1}     $ & $ 170.8^{+  14.1}_{ -13.1}     $ & $  63.6^{+   9.1}_{  -8.0}     $ & $  16.6^{+   5.2}_{  -4.1}     $ & $ 21.00^{+  1.50}_{ -1.40}     $ & $  6.30^{+  0.52}_{ -0.48}     $ & $ 13.00^{+  1.80}_{ -1.60}     $ & $  6.50^{+  2.00}_{ -1.60}     $ &$10^{ -8.0}          $ &   1.30 &  -0.4 & fshu \\
c61   & J141735.9+523029 & $2674.1^{+  52.7}_{ -51.7}     $ & $1951.5^{+  45.2}_{ -44.2}     $ & $ 760.3^{+  28.6}_{ -27.6}     $ & $ 239.3^{+  16.5}_{ -15.5}     $ & $160.00^{+  3.20}_{ -3.20}     $ & $ 47.00^{+  1.10}_{ -1.10}     $ & $100.00^{+  3.80}_{ -3.70}     $ & $ 62.00^{+  4.20}_{ -4.00}     $ &$10^{ -8.0}          $ &   2.30 &  -0.4 & fshu \\
c62   & J141736.3+523016 & $   9.9^{+   4.4}_{  -3.3}     $ & $   6.5^{+   3.8}_{  -2.6}     $ & $   3.3^{+   3.2}_{  -1.9}     $ & $<  4.34                       $ & $  0.60^{+  0.24}_{ -0.18}     $ & $  0.16^{+  0.09}_{ -0.06}     $ & $  0.44^{+  0.35}_{ -0.21}     $ & $<  1.10                       $ &$10^{ -6.4}          $ &   2.10 &  -0.3 & f    \\
c63   & J141736.3+523544 & $  53.4^{+   9.1}_{  -8.0}     $ & $  43.5^{+   8.1}_{  -7.0}     $ & $  12.2^{+   5.6}_{  -4.4}     $ & $<  9.22                       $ & $  4.00^{+  0.56}_{ -0.49}     $ & $  1.30^{+  0.21}_{ -0.18}     $ & $  2.10^{+  0.57}_{ -0.45}     $ & $<  3.10                       $ &$10^{ -8.0}          $ &   7.40 &  -0.6 & fs   \\
c64   & J141736.8+522429 & $  93.8^{+  10.8}_{  -9.8}     $ & $  69.0^{+   9.4}_{  -8.3}     $ & $  26.4^{+   6.4}_{  -5.3}     $ & $   7.9^{+   4.1}_{  -2.9}     $ & $  6.00^{+  0.68}_{ -0.61}     $ & $  1.80^{+  0.24}_{ -0.21}     $ & $  3.70^{+  0.84}_{ -0.70}     $ & $  2.20^{+  0.99}_{ -0.71}     $ &$10^{ -8.0}          $ &   4.00 &  -0.4 & fshu \\
c65   & J141737.3+522921 & $  13.4^{+   4.8}_{  -3.7}     $ & $<  2.79                       $ & $  13.6^{+   4.8}_{  -3.7}     $ & $   4.6^{+   3.4}_{  -2.2}     $ & $  1.10^{+  0.37}_{ -0.28}     $ & $<  0.09                       $ & $  2.40^{+  0.82}_{ -0.62}     $ & $  1.50^{+  1.00}_{ -0.66}     $ &$10^{ -8.0}          $ &   1.30 &   1.0 & fhu  \\
c66   & J141738.7+523413 & $  23.0^{+   6.4}_{  -5.3}     $ & $  16.4^{+   5.4}_{  -4.3}     $ & $   6.7^{+   4.4}_{  -3.3}     $ & $<  7.32                       $ & $  1.50^{+  0.34}_{ -0.28}     $ & $  0.42^{+  0.12}_{ -0.10}     $ & $  0.98^{+  0.39}_{ -0.29}     $ & $<  2.10                       $ &$10^{ -8.0}          $ &   5.80 &  -0.4 & fs   \\
c67   & J141738.8+522332 & $ 311.6^{+  18.8}_{ -17.7}     $ & $ 232.4^{+  16.3}_{ -15.3}     $ & $  87.6^{+  10.5}_{  -9.5}     $ & $  27.4^{+   6.4}_{  -5.3}     $ & $ 20.00^{+  1.20}_{ -1.20}     $ & $  6.00^{+  0.42}_{ -0.39}     $ & $ 13.00^{+  1.50}_{ -1.30}     $ & $  7.70^{+  1.70}_{ -1.40}     $ &$10^{ -8.0}          $ &   4.90 &  -0.4 & fshu \\
c68   & J141739.0+522843 & $   4.7^{+   3.4}_{  -2.2}     $ & $   3.8^{+   3.2}_{  -1.9}     $ & $<  3.76                       $ & $<  3.78                       $ & $  0.66^{+  0.45}_{ -0.29}     $ & $  0.21^{+  0.17}_{ -0.10}     $ & $<  1.10                       $ & $<  2.20                       $ &$10^{ -6.5}          $ &   0.70 &  -1.0 & fs   \\
c69   & J141739.3+522850 & $   4.7^{+   3.4}_{  -2.2}     $ & $<  2.86                       $ & $   4.8^{+   3.4}_{  -2.2}     $ & $   3.8^{+   3.2}_{  -1.9}     $ & $  0.77^{+  0.52}_{ -0.33}     $ & $<  0.18                       $ & $  1.70^{+  1.20}_{ -0.75}     $ & $  2.70^{+  2.10}_{ -1.30}     $ &$10^{ -7.2}          $ &   0.70 &   1.0 & hf   \\
c70   & J141739.5+523619 & $  49.5^{+   9.1}_{  -8.0}     $ & $  35.0^{+   7.5}_{  -6.4}     $ & $  12.5^{+   5.9}_{  -4.8}     $ & $<  9.57                       $ & $  3.40^{+  0.47}_{ -0.42}     $ & $  0.94^{+  0.17}_{ -0.14}     $ & $  1.90^{+  0.49}_{ -0.40}     $ & $<  2.90                       $ &$10^{ -8.0}          $ &   7.90 &  -0.5 & fs   \\
c71   & J141741.4+523545 & $  30.9^{+   7.7}_{  -6.6}     $ & $  10.9^{+   5.2}_{  -4.1}     $ & $  19.3^{+   6.4}_{  -5.3}     $ & $<  9.62                       $ & $  2.10^{+  0.36}_{ -0.31}     $ & $  0.29^{+  0.09}_{ -0.07}     $ & $  2.90^{+  0.66}_{ -0.55}     $ & $<  2.90                       $ &$10^{ -8.0}          $ &   7.30 &   0.3 & fsh  \\
c72   & J141741.9+522823 & $ 817.2^{+  29.6}_{ -28.6}     $ & $ 544.7^{+  24.4}_{ -23.3}     $ & $ 274.5^{+  17.6}_{ -16.6}     $ & $  72.5^{+   9.6}_{  -8.5}     $ & $ 50.00^{+  1.80}_{ -1.80}     $ & $ 13.00^{+  0.59}_{ -0.56}     $ & $ 38.00^{+  2.40}_{ -2.30}     $ & $ 20.00^{+  2.60}_{ -2.30}     $ &$10^{ -8.0}          $ &   0.20 &  -0.3 & fshu \\
c73   & J141742.8+522235 & $   3.7^{+   3.6}_{  -2.4}     $ & $   4.9^{+   3.6}_{  -2.4}     $ & $<  6.40                       $ & $<  4.91                       $ & $  0.63^{+  0.38}_{ -0.25}     $ & $  0.34^{+  0.20}_{ -0.13}     $ & $<  2.30                       $ & $<  3.60                       $ &$10^{ -5.7}          $ &   5.80 &  -1.0 & s    \\
c74   & J141743.2+522023 & $  14.3^{+   5.7}_{  -4.5}     $ & $  12.1^{+   5.0}_{  -3.8}     $ & $<  8.45                       $ & $<  6.50                       $ & $  2.20^{+  0.58}_{ -0.46}     $ & $  0.69^{+  0.23}_{ -0.18}     $ & $<  2.60                       $ & $<  4.20                       $ &$10^{ -6.9}          $ &   8.00 &  -1.0 & sf   \\
c75   & J141745.4+522951 & $  52.2^{+   8.3}_{  -7.2}     $ & $   7.7^{+   4.0}_{  -2.8}     $ & $  45.4^{+   7.8}_{  -6.8}     $ & $  18.5^{+   5.4}_{  -4.3}     $ & $  3.20^{+  0.51}_{ -0.44}     $ & $  0.19^{+  0.09}_{ -0.06}     $ & $  6.30^{+  1.10}_{ -0.93}     $ & $  5.10^{+  1.50}_{ -1.20}     $ &$10^{ -8.0}          $ &   1.50 &   0.7 & fshu \\
c76   & J141745.7+522801 & $  54.6^{+   8.5}_{  -7.4}     $ & $  37.8^{+   7.2}_{  -6.1}     $ & $  17.6^{+   5.3}_{  -4.2}     $ & $   7.7^{+   4.0}_{  -2.8}     $ & $  6.10^{+  0.94}_{ -0.82}     $ & $  1.70^{+  0.31}_{ -0.27}     $ & $  4.40^{+  1.30}_{ -1.00}     $ & $  3.80^{+  1.90}_{ -1.30}     $ &$10^{ -8.0}          $ &   0.60 &  -0.3 & fshu \\
c77   & J141745.9+523032 & $ 239.2^{+  16.5}_{ -15.5}     $ & $ 121.7^{+  12.1}_{ -11.0}     $ & $ 119.4^{+  12.0}_{ -10.9}     $ & $  38.4^{+   7.3}_{  -6.2}     $ & $ 15.00^{+  1.00}_{ -0.96}     $ & $  3.00^{+  0.29}_{ -0.27}     $ & $ 17.00^{+  1.70}_{ -1.50}     $ & $ 10.00^{+  2.00}_{ -1.70}     $ &$10^{ -8.0}          $ &   2.20 &   0.0 & fshu \\
c78   & J141746.1+522526 & $  10.9^{+   4.6}_{  -3.4}     $ & $   3.4^{+   3.2}_{  -1.9}     $ & $   7.1^{+   4.0}_{  -2.8}     $ & $<  4.32                       $ & $  0.68^{+  0.26}_{ -0.19}     $ & $  0.08^{+  0.07}_{ -0.04}     $ & $  0.99^{+  0.49}_{ -0.34}     $ & $<  1.10                       $ &$10^{ -6.4}          $ &   3.00 &   0.4 & f    \\
\hline
\end{tabular}
\end{center}
\end{table}
\end{landscape}

\setcounter{table}{2}
\begin{landscape}
\begin{table}
\centering
\caption{Chandra GWS X-ray catalogue (continued)}
\begin{center}
\begin{tabular}{@{}ccrrrrrrrrcccc@{}}
\hline
Cat & CXO GWS & FB & SB & HB & UB & 0.5-10.0 & 0.5-2.0 & 2-10 & 5-10 & 
  $p_{\rm min}$ & OAA & HR & Flags \\
No. & (2000) & cts & cts & cts & cts & flux & flux & flux & flux & & ($\prime$) & &  \\
(1) & (2) & (3) & (4) & (5) & (6) & (7) & (8) & (9) & (10) & (11) & (12) & (13) & (14) \\
\hline
c79   & J141746.7+522858 & $  12.3^{+   4.7}_{  -3.5}     $ & $   8.7^{+   4.1}_{  -2.9}     $ & $   3.4^{+   3.2}_{  -1.9}     $ & $<  4.46                       $ & $  0.76^{+  0.28}_{ -0.21}     $ & $  0.21^{+  0.10}_{ -0.07}     $ & $  0.47^{+  0.38}_{ -0.23}     $ & $<  1.20                       $ &$10^{ -8.0}          $ &   0.80 &  -0.4 & fs   \\
c80   & J141747.0+522512 & $   9.5^{+   4.4}_{  -3.3}     $ & $<  4.28                       $ & $   6.9^{+   4.0}_{  -2.8}     $ & $<  5.14                       $ & $  0.60^{+  0.24}_{ -0.18}     $ & $<  0.11                       $ & $  0.96^{+  0.48}_{ -0.33}     $ & $<  1.30                       $ &$10^{ -7.2}          $ &   3.30 &   1.0 & hf   \\
c81   & J141747.0+522816 & $  11.6^{+   4.6}_{  -3.4}     $ & $   8.8^{+   4.1}_{  -2.9}     $ & $   5.7^{+   3.6}_{  -2.4}     $ & $<  3.73                       $ & $  1.40^{+  0.55}_{ -0.41}     $ & $  0.43^{+  0.20}_{ -0.14}     $ & $  1.60^{+  0.96}_{ -0.63}     $ & $<  2.00                       $ &$10^{ -8.0}          $ &   0.60 &  -0.2 & fsh  \\
c82   & J141747.4+523510 & $  87.0^{+  10.9}_{  -9.8}     $ & $  68.3^{+   9.6}_{  -8.5}     $ & $  26.5^{+   6.8}_{  -5.7}     $ & $<  7.87                       $ & $  5.80^{+  0.65}_{ -0.59}     $ & $  1.80^{+  0.24}_{ -0.21}     $ & $  4.00^{+  0.82}_{ -0.69}     $ & $<  2.30                       $ &$10^{ -8.0}          $ &   6.80 &  -0.4 & fsh  \\
c83   & J141749.2+522803 & $  10.6^{+   4.4}_{  -3.3}     $ & $   3.8^{+   3.2}_{  -1.9}     $ & $   6.7^{+   3.8}_{  -2.6}     $ & $<  3.71                       $ & $  1.10^{+  0.45}_{ -0.33}     $ & $  0.16^{+  0.13}_{ -0.08}     $ & $  1.60^{+  0.85}_{ -0.58}     $ & $<  1.60                       $ &$10^{ -8.0}          $ &   1.00 &   0.3 & fh   \\
c84   & J141749.2+522811 & $  58.7^{+   8.7}_{  -7.7}     $ & $  35.9^{+   7.1}_{  -6.0}     $ & $  22.8^{+   5.9}_{  -4.8}     $ & $   3.8^{+   3.2}_{  -1.9}     $ & $  9.00^{+  1.30}_{ -1.20}     $ & $  2.20^{+  0.42}_{ -0.36}     $ & $  7.90^{+  2.00}_{ -1.60}     $ & $  2.60^{+  2.10}_{ -1.20}     $ &$10^{ -8.0}          $ &   1.00 &  -0.2 & fsh  \\
c85   & J141749.7+523143 & $  48.5^{+   8.1}_{  -7.0}     $ & $  26.3^{+   6.3}_{  -5.2}     $ & $  22.9^{+   6.0}_{  -4.9}     $ & $  11.1^{+   4.6}_{  -3.4}     $ & $  3.10^{+  0.50}_{ -0.43}     $ & $  0.65^{+  0.15}_{ -0.13}     $ & $  3.30^{+  0.81}_{ -0.66}     $ & $  3.20^{+  1.20}_{ -0.90}     $ &$10^{ -8.0}          $ &   3.50 &  -0.1 & fshu \\
c86   & J141750.1+523601 & $  39.3^{+   8.3}_{  -7.2}     $ & $  30.0^{+   7.1}_{  -6.0}     $ & $  11.4^{+   5.7}_{  -4.5}     $ & $< 10.02                       $ & $  2.80^{+  0.43}_{ -0.38}     $ & $  0.82^{+  0.16}_{ -0.14}     $ & $  1.80^{+  0.48}_{ -0.38}     $ & $<  3.10                       $ &$10^{ -8.0}          $ &   7.70 &  -0.4 & fs   \\
c87   & J141750.5+522339 & $  25.2^{+   6.4}_{  -5.3}     $ & $  10.6^{+   4.6}_{  -3.4}     $ & $  15.1^{+   5.2}_{  -4.1}     $ & $   3.7^{+   3.4}_{  -2.2}     $ & $  2.10^{+  0.48}_{ -0.39}     $ & $  0.35^{+  0.13}_{ -0.10}     $ & $  2.80^{+  0.85}_{ -0.67}     $ & $  1.30^{+  0.90}_{ -0.57}     $ &$10^{ -8.0}          $ &   4.90 &   0.2 & fsh  \\
c88   & J141750.8+523632 & $  23.1^{+   7.3}_{  -6.2}     $ & $  16.5^{+   5.9}_{  -4.8}     $ & $< 11.80                       $ & $<  9.98                       $ & $  1.70^{+  0.31}_{ -0.27}     $ & $  0.48^{+  0.12}_{ -0.10}     $ & $<  1.80                       $ & $<  3.20                       $ &$10^{ -8.0}          $ &   8.20 &  -1.0 & fs   \\
c89   & J141751.0+522534 & $  24.0^{+   6.1}_{  -5.0}     $ & $<  4.50                       $ & $  23.2^{+   6.0}_{  -4.9}     $ & $  12.4^{+   4.7}_{  -3.5}     $ & $  1.70^{+  0.41}_{ -0.33}     $ & $<  0.12                       $ & $  3.60^{+  0.90}_{ -0.73}     $ & $  3.70^{+  1.40}_{ -1.00}     $ &$10^{ -8.0}          $ &   3.10 &   1.0 & fhu  \\
c90   & J141751.1+522310 & $ 127.1^{+  12.5}_{ -11.4}     $ & $  85.1^{+  10.4}_{  -9.3}     $ & $  43.4^{+   7.8}_{  -6.8}     $ & $  15.2^{+   5.2}_{  -4.1}     $ & $  9.90^{+  0.94}_{ -0.86}     $ & $  2.60^{+  0.31}_{ -0.28}     $ & $  7.50^{+  1.30}_{ -1.10}     $ & $  5.20^{+  1.60}_{ -1.30}     $ &$10^{ -8.0}          $ &   5.40 &  -0.3 & fshu \\
c91   & J141751.7+523046 & $  16.1^{+   5.2}_{  -4.1}     $ & $  11.6^{+   4.6}_{  -3.4}     $ & $   4.2^{+   3.4}_{  -2.2}     $ & $<  4.37                       $ & $  1.10^{+  0.33}_{ -0.26}     $ & $  0.30^{+  0.11}_{ -0.09}     $ & $  0.63^{+  0.43}_{ -0.27}     $ & $<  1.20                       $ &$10^{ -8.0}          $ &   2.70 &  -0.4 & fs   \\
c92   & J141752.4+522853 & $  26.3^{+   6.3}_{  -5.2}     $ & $<  3.70                       $ & $  24.4^{+   6.1}_{  -5.0}     $ & $   8.5^{+   4.1}_{  -2.9}     $ & $  1.70^{+  0.40}_{ -0.33}     $ & $<  0.09                       $ & $  3.60^{+  0.88}_{ -0.71}     $ & $  2.50^{+  1.10}_{ -0.81}     $ &$10^{ -8.0}          $ &   1.50 &   1.0 & fhu  \\
c93   & J141752.9+522838 & $   6.3^{+   3.8}_{  -2.6}     $ & $<  3.70                       $ & $<  4.45                       $ & $<  4.50                       $ & $  0.42^{+  0.23}_{ -0.15}     $ & $<  0.10                       $ & $<  0.63                       $ & $<  1.30                       $ &$10^{ -6.0}          $ &   1.50 & ... & f    \\
c94   & J141753.1+522050 & $  21.0^{+   6.9}_{  -5.8}     $ & $  10.8^{+   5.2}_{  -4.1}     $ & $  10.4^{+   5.4}_{  -4.3}     $ & $   3.7^{+   4.1}_{  -2.9}     $ & $  1.60^{+  0.32}_{ -0.27}     $ & $  0.32^{+  0.10}_{ -0.08}     $ & $  1.80^{+  0.50}_{ -0.40}     $ & $  1.30^{+  0.59}_{ -0.42}     $ &$10^{ -7.2}          $ &   7.70 &   0.0 & f    \\
c95   & J141753.7+523446 & $  53.7^{+   9.0}_{  -7.9}     $ & $<  9.07                       $ & $  47.2^{+   8.3}_{  -7.2}     $ & $  24.6^{+   6.4}_{  -5.3}     $ & $  3.60^{+  0.51}_{ -0.45}     $ & $<  0.24                       $ & $  7.00^{+  1.10}_{ -0.96}     $ & $  7.50^{+  1.70}_{ -1.40}     $ &$10^{ -8.0}          $ &   6.60 &   1.0 & fhu  \\
c96   & J141753.9+523033 & $  17.1^{+   5.3}_{  -4.2}     $ & $  10.6^{+   4.4}_{  -3.3}     $ & $   6.3^{+   3.8}_{  -2.6}     $ & $<  4.35                       $ & $  1.10^{+  0.32}_{ -0.25}     $ & $  0.26^{+  0.11}_{ -0.08}     $ & $  0.89^{+  0.48}_{ -0.33}     $ & $<  1.10                       $ &$10^{ -8.0}          $ &   2.70 &  -0.2 & fsh  \\
c97   & J141754.2+523123 & $  50.6^{+   8.3}_{  -7.2}     $ & $  40.4^{+   7.5}_{  -6.4}     $ & $  11.9^{+   4.7}_{  -3.5}     $ & $<  5.08                       $ & $  3.20^{+  0.51}_{ -0.44}     $ & $  1.00^{+  0.18}_{ -0.16}     $ & $  1.70^{+  0.61}_{ -0.46}     $ & $<  1.40                       $ &$10^{ -8.0}          $ &   3.40 &  -0.5 & fsh  \\
c98   & J141754.5+523437 & $  15.2^{+   6.0}_{  -4.9}     $ & $<  8.36                       $ & $  15.4^{+   5.7}_{  -4.5}     $ & $<  3.96                       $ & $  1.00^{+  0.25}_{ -0.20}     $ & $<  0.22                       $ & $  2.30^{+  0.62}_{ -0.50}     $ & $<  0.72                       $ &$10^{ -8.0}          $ &   6.50 &   1.0 & hf   \\
c99   & J141755.2+523532 & $  40.3^{+   8.3}_{  -7.2}     $ & $  29.5^{+   7.1}_{  -6.0}     $ & $  15.9^{+   6.0}_{  -4.9}     $ & $<  8.33                       $ & $  2.70^{+  0.43}_{ -0.37}     $ & $  0.79^{+  0.15}_{ -0.13}     $ & $  2.40^{+  0.60}_{ -0.49}     $ & $<  2.50                       $ &$10^{ -8.0}          $ &   7.40 &  -0.3 & fsh  \\
c100  & J141756.7+522400 & $  38.7^{+   7.5}_{  -6.4}     $ & $  25.4^{+   6.3}_{  -5.2}     $ & $  14.8^{+   5.2}_{  -4.1}     $ & $<  5.45                       $ & $  2.60^{+  0.47}_{ -0.40}     $ & $  0.68^{+  0.16}_{ -0.13}     $ & $  2.20^{+  0.69}_{ -0.54}     $ & $<  1.50                       $ &$10^{ -8.0}          $ &   4.90 &  -0.2 & fsh  \\
c101  & J141756.8+523124 & $  72.2^{+   9.6}_{  -8.6}     $ & $  45.2^{+   7.8}_{  -6.8}     $ & $  27.8^{+   6.4}_{  -5.3}     $ & $   9.0^{+   4.3}_{  -3.1}     $ & $  4.60^{+  0.60}_{ -0.54}     $ & $  1.10^{+  0.19}_{ -0.17}     $ & $  4.00^{+  0.88}_{ -0.73}     $ & $  2.60^{+  1.10}_{ -0.79}     $ &$10^{ -8.0}          $ &   3.70 &  -0.2 & fshu \\
c102  & J141756.9+523118 & $   5.3^{+   3.8}_{  -2.6}     $ & $   4.2^{+   3.4}_{  -2.2}     $ & $<  5.76                       $ & $<  4.99                       $ & $  0.34^{+  0.18}_{ -0.13}     $ & $  0.11^{+  0.07}_{ -0.04}     $ & $<  0.78                       $ & $<  1.30                       $ &$10^{ -5.4}          $ &   3.60 &  -1.0 & f    \\
c103  & J141757.1+522630 & $ 100.0^{+  11.1}_{ -10.0}     $ & $  46.5^{+   7.9}_{  -6.8}     $ & $  59.1^{+   8.8}_{  -7.7}     $ & $  15.3^{+   5.1}_{  -4.0}     $ & $  6.30^{+  0.69}_{ -0.62}     $ & $  1.10^{+  0.19}_{ -0.17}     $ & $  8.20^{+  1.20}_{ -1.10}     $ & $  4.10^{+  1.30}_{ -1.00}     $ &$10^{ -8.0}          $ &   2.90 &   0.1 & fshu \\
c104  & J141757.4+523106 & $  29.4^{+   6.6}_{  -5.5}     $ & $  16.2^{+   5.2}_{  -4.1}     $ & $  12.7^{+   4.8}_{  -3.7}     $ & $   8.1^{+   4.1}_{  -2.9}     $ & $  1.90^{+  0.40}_{ -0.34}     $ & $  0.41^{+  0.12}_{ -0.10}     $ & $  1.80^{+  0.63}_{ -0.48}     $ & $  2.30^{+  1.00}_{ -0.74}     $ &$10^{ -8.0}          $ &   3.50 &  -0.1 & fshu \\
c105  & J141757.5+522546 & $  44.5^{+   7.8}_{  -6.8}     $ & $  28.2^{+   6.4}_{  -5.3}     $ & $  16.9^{+   5.3}_{  -4.2}     $ & $<  4.21                       $ & $  2.80^{+  0.48}_{ -0.41}     $ & $  0.70^{+  0.16}_{ -0.13}     $ & $  2.40^{+  0.70}_{ -0.55}     $ & $<  1.10                       $ &$10^{ -8.0}          $ &   3.40 &  -0.2 & fsh  \\
c106  & J141758.1+523133 & $   9.9^{+   4.6}_{  -3.4}     $ & $   7.1^{+   4.0}_{  -2.8}     $ & $<  5.56                       $ & $<  4.95                       $ & $  0.63^{+  0.24}_{ -0.18}     $ & $  0.18^{+  0.09}_{ -0.06}     $ & $<  0.75                       $ & $<  1.30                       $ &$10^{ -7.2}          $ &   3.90 &  -1.0 & f    \\
c107  & J141758.2+522153 & $  54.3^{+   9.1}_{  -8.0}     $ & $  36.4^{+   7.5}_{  -6.4}     $ & $  20.5^{+   6.3}_{  -5.2}     $ & $<  8.74                       $ & $  3.90^{+  0.55}_{ -0.48}     $ & $  1.00^{+  0.18}_{ -0.16}     $ & $  3.30^{+  0.76}_{ -0.63}     $ & $<  2.70                       $ &$10^{ -8.0}          $ &   6.90 &  -0.3 & fsh  \\
c108  & J141758.9+523138 & $ 137.9^{+  12.9}_{ -11.8}     $ & $  84.0^{+  10.3}_{  -9.2}     $ & $  53.5^{+   8.5}_{  -7.4}     $ & $  13.9^{+   5.0}_{  -3.8}     $ & $  8.90^{+  0.82}_{ -0.75}     $ & $  2.10^{+  0.26}_{ -0.23}     $ & $  7.70^{+  1.20}_{ -1.00}     $ & $  4.00^{+  1.30}_{ -1.00}     $ &$10^{ -8.0}          $ &   4.00 &  -0.2 & fshu \\
c109  & J141759.3+522420 & $  19.8^{+   5.9}_{  -4.8}     $ & $  16.5^{+   5.3}_{  -4.2}     $ & $   3.8^{+   3.6}_{  -2.4}     $ & $<  5.57                       $ & $  1.30^{+  0.33}_{ -0.26}     $ & $  0.42^{+  0.12}_{ -0.10}     $ & $  0.55^{+  0.33}_{ -0.22}     $ & $<  1.50                       $ &$10^{ -8.0}          $ &   4.80 &  -0.6 & fs   \\
c110  & J141800.0+522223 & $ 162.1^{+  14.1}_{ -13.1}     $ & $  80.0^{+  10.2}_{  -9.1}     $ & $  86.8^{+  10.7}_{  -9.6}     $ & $  30.2^{+   6.9}_{  -5.8}     $ & $ 11.00^{+  0.90}_{ -0.83}     $ & $  2.10^{+  0.26}_{ -0.23}     $ & $ 13.00^{+  1.50}_{ -1.30}     $ & $  9.20^{+  1.90}_{ -1.60}     $ &$10^{ -8.0}          $ &   6.60 &   0.1 & fshu \\
c111  & J141800.4+522822 & $   9.5^{+   4.3}_{  -3.1}     $ & $<  2.94                       $ & $   7.6^{+   4.0}_{  -2.8}     $ & $<  3.69                       $ & $  1.30^{+  0.56}_{ -0.40}     $ & $<  0.15                       $ & $  2.40^{+  1.20}_{ -0.82}     $ & $<  2.20                       $ &$10^{ -8.0}          $ &   2.60 &   0.5 & fh   \\
c112  & J141800.4+523610 & $  64.2^{+  10.0}_{  -9.0}     $ & $  49.3^{+   8.7}_{  -7.6}     $ & $  14.9^{+   6.2}_{  -5.1}     $ & $   5.1^{+   4.6}_{  -3.4}     $ & $  4.40^{+  0.55}_{ -0.49}     $ & $  1.30^{+  0.20}_{ -0.18}     $ & $  2.30^{+  0.55}_{ -0.45}     $ & $  1.60^{+  0.63}_{ -0.47}     $ &$10^{ -8.0}          $ &   8.20 &  -0.5 & fsh  \\
c113  & J141801.1+522941 & $  18.9^{+   5.6}_{  -4.4}     $ & $  17.4^{+   5.3}_{  -4.2}     $ & $<  5.07                       $ & $<  4.28                       $ & $  1.20^{+  0.34}_{ -0.27}     $ & $  0.44^{+  0.13}_{ -0.10}     $ & $<  0.69                       $ & $<  1.20                       $ &$10^{ -8.0}          $ &   3.00 &  -1.0 & fs   \\
c114  & J141801.3+523150 & $  11.5^{+   4.8}_{  -3.7}     $ & $<  5.76                       $ & $   9.1^{+   4.4}_{  -3.3}     $ & $<  5.76                       $ & $  0.74^{+  0.26}_{ -0.20}     $ & $<  0.15                       $ & $  1.30^{+  0.53}_{ -0.39}     $ & $<  1.60                       $ &$10^{ -8.0}          $ &   4.40 &   1.0 & fh   \\
c115  & J141801.6+522800 & $  11.9^{+   4.7}_{  -3.5}     $ & $   8.4^{+   4.1}_{  -2.9}     $ & $<  5.09                       $ & $<  4.31                       $ & $  0.75^{+  0.27}_{ -0.20}     $ & $  0.21^{+  0.10}_{ -0.07}     $ & $<  0.68                       $ & $<  1.20                       $ &$10^{ -8.0}          $ &   2.90 &  -1.0 & fs   \\
c116  & J141802.0+523514 & $ 324.3^{+  19.4}_{ -18.4}     $ & $ 253.5^{+  17.1}_{ -16.1}     $ & $  80.5^{+  10.5}_{  -9.4}     $ & $  24.8^{+   6.6}_{  -5.4}     $ & $ 22.00^{+  1.30}_{ -1.20}     $ & $  6.90^{+  0.45}_{ -0.43}     $ & $ 12.00^{+  1.50}_{ -1.30}     $ & $  7.90^{+  1.70}_{ -1.40}     $ &$10^{ -8.0}          $ &   7.40 &  -0.5 & fshu \\
c117  & J141802.4+522132 & $  53.5^{+   9.2}_{  -8.1}     $ & $  40.6^{+   7.9}_{  -6.8}     $ & $  14.9^{+   5.9}_{  -4.8}     $ & $   5.9^{+   4.4}_{  -3.3}     $ & $  3.70^{+  0.51}_{ -0.45}     $ & $  1.10^{+  0.18}_{ -0.16}     $ & $  2.30^{+  0.59}_{ -0.47}     $ & $  1.90^{+  0.75}_{ -0.55}     $ &$10^{ -8.0}          $ &   7.50 &  -0.4 & fsh  \\
c118  & J141802.9+523547 & $ 100.1^{+  11.8}_{ -10.7}     $ & $  71.3^{+   9.9}_{  -8.9}     $ & $  29.6^{+   7.3}_{  -6.2}     $ & $   9.3^{+   5.0}_{  -3.8}     $ & $  7.50^{+  0.76}_{ -0.69}     $ & $  2.10^{+  0.26}_{ -0.24}     $ & $  4.90^{+  0.92}_{ -0.78}     $ & $  3.20^{+  1.10}_{ -0.82}     $ &$10^{ -8.0}          $ &   8.00 &  -0.4 & fshu \\
\hline
\end{tabular}
\end{center}
\end{table}
\end{landscape}

\setcounter{table}{2}
\begin{landscape}
\begin{table}
\centering
\caption{Chandra GWS X-ray catalogue (continued)}
\begin{center}
\begin{tabular}{@{}ccrrrrrrrrcccc@{}}
\hline
Cat & CXO GWS & FB & SB & HB & UB & 0.5-10.0 & 0.5-2.0 & 2-10 & 5-10 & 
  $p_{\rm min}$ & OAA & HR & Flags \\
No. & (2000) & cts & cts & cts & cts & flux & flux & flux & flux & & ($\prime$) & &  \\
(1) & (2) & (3) & (4) & (5) & (6) & (7) & (8) & (9) & (10) & (11) & (12) & (13) & (14) \\
\hline
c119  & J141804.5+523633 & $ 150.5^{+  14.0}_{ -13.0}     $ & $ 119.8^{+  12.4}_{ -11.3}     $ & $  42.0^{+   8.4}_{  -7.3}     $ & $  10.9^{+   5.3}_{  -4.2}     $ & $ 12.00^{+  1.00}_{ -0.92}     $ & $  3.80^{+  0.36}_{ -0.33}     $ & $  7.50^{+  1.20}_{ -1.00}     $ & $  4.10^{+  1.20}_{ -0.96}     $ &$10^{ -8.0}          $ &   8.80 &  -0.5 & fshu \\
c120  & J141804.8+522740 & $  44.4^{+   7.8}_{  -6.8}     $ & $  31.2^{+   6.7}_{  -5.6}     $ & $  13.8^{+   5.0}_{  -3.8}     $ & $<  5.07                       $ & $  2.90^{+  0.50}_{ -0.43}     $ & $  0.81^{+  0.17}_{ -0.14}     $ & $  2.00^{+  0.67}_{ -0.52}     $ & $<  1.40                       $ &$10^{ -8.0}          $ &   3.40 &  -0.4 & fsh  \\
c121  & J141805.3+522510 & $  38.0^{+   7.5}_{  -6.4}     $ & $  24.6^{+   6.2}_{  -5.1}     $ & $  13.9^{+   5.1}_{  -4.0}     $ & $<  5.62                       $ & $  2.60^{+  0.48}_{ -0.41}     $ & $  0.67^{+  0.16}_{ -0.13}     $ & $  2.10^{+  0.68}_{ -0.53}     $ & $<  1.60                       $ &$10^{ -8.0}          $ &   4.70 &  -0.3 & fsh  \\
c122  & J141806.5+523358 & $  19.4^{+   6.3}_{  -5.2}     $ & $  11.6^{+   5.0}_{  -3.8}     $ & $   6.9^{+   4.6}_{  -3.4}     $ & $<  7.79                       $ & $  1.40^{+  0.33}_{ -0.27}     $ & $  0.34^{+  0.11}_{ -0.09}     $ & $  1.10^{+  0.43}_{ -0.32}     $ & $<  2.50                       $ &$10^{ -8.0}          $ &   6.60 &  -0.2 & fs   \\
c123  & J141807.0+522523 & $ 102.6^{+  11.3}_{ -10.3}     $ & $  59.3^{+   8.9}_{  -7.8}     $ & $  44.8^{+   7.9}_{  -6.8}     $ & $  11.4^{+   4.7}_{  -3.5}     $ & $  6.60^{+  0.71}_{ -0.65}     $ & $  1.50^{+  0.22}_{ -0.19}     $ & $  6.50^{+  1.10}_{ -0.94}     $ & $  3.20^{+  1.20}_{ -0.89}     $ &$10^{ -8.0}          $ &   4.70 &  -0.1 & fsh  \\
c124  & J141807.3+523030 & $   4.9^{+   3.8}_{  -2.6}     $ & $   4.1^{+   3.4}_{  -2.2}     $ & $<  5.40                       $ & $<  5.81                       $ & $  0.37^{+  0.20}_{ -0.13}     $ & $  0.12^{+  0.08}_{ -0.05}     $ & $<  0.85                       $ & $<  1.90                       $ &$10^{ -5.6}          $ &   4.20 &  -1.0 & s    \\
c125  & J141808.0+522750 & $  11.9^{+   4.8}_{  -3.7}     $ & $   5.9^{+   3.8}_{  -2.6}     $ & $   6.5^{+   4.0}_{  -2.8}     $ & $   6.0^{+   3.8}_{  -2.6}     $ & $  0.79^{+  0.27}_{ -0.21}     $ & $  0.15^{+  0.08}_{ -0.06}     $ & $  0.97^{+  0.48}_{ -0.34}     $ & $  1.80^{+  0.97}_{ -0.66}     $ &$10^{ -6.6}          $ &   3.90 &   0.1 & f    \\
c126  & J141808.9+523150 & $  10.0^{+   4.8}_{  -3.7}     $ & $<  6.14                       $ & $   8.1^{+   4.4}_{  -3.3}     $ & $   7.0^{+   4.1}_{  -2.9}     $ & $  0.66^{+  0.23}_{ -0.17}     $ & $<  0.16                       $ & $  1.20^{+  0.48}_{ -0.35}     $ & $  2.00^{+  0.94}_{ -0.67}     $ &$10^{ -6.4}          $ &   5.20 &   1.0 & f    \\
c127  & J141809.1+522804 & $  49.7^{+   8.3}_{  -7.2}     $ & $  24.9^{+   6.2}_{  -5.1}     $ & $  24.4^{+   6.2}_{  -5.1}     $ & $  10.0^{+   4.4}_{  -3.3}     $ & $  3.50^{+  0.56}_{ -0.49}     $ & $  0.70^{+  0.17}_{ -0.14}     $ & $  4.00^{+  0.94}_{ -0.77}     $ & $  3.30^{+  1.30}_{ -0.97}     $ &$10^{ -8.0}          $ &   4.00 &   0.0 & fshu \\
c128  & J141811.2+523011 & $  15.6^{+   5.4}_{  -4.3}     $ & $   9.6^{+   4.4}_{  -3.3}     $ & $   6.6^{+   4.1}_{  -2.9}     $ & $<  6.33                       $ & $  1.00^{+  0.29}_{ -0.23}     $ & $  0.25^{+  0.10}_{ -0.07}     $ & $  0.97^{+  0.44}_{ -0.32}     $ & $<  1.80                       $ &$10^{ -8.0}          $ &   4.60 &  -0.2 & fs   \\
c129  & J141812.1+522800 & $  29.3^{+   6.8}_{  -5.7}     $ & $  20.3^{+   5.8}_{  -4.7}     $ & $   9.6^{+   4.6}_{  -3.4}     $ & $<  5.46                       $ & $  1.90^{+  0.39}_{ -0.33}     $ & $  0.52^{+  0.14}_{ -0.11}     $ & $  1.40^{+  0.54}_{ -0.40}     $ & $<  1.50                       $ &$10^{ -8.0}          $ &   4.40 &  -0.3 & fs   \\
c130  & J141813.1+523113 & $  50.4^{+   8.5}_{  -7.4}     $ & $  34.9^{+   7.1}_{  -6.1}     $ & $  17.9^{+   5.7}_{  -4.5}     $ & $   8.8^{+   4.4}_{  -3.3}     $ & $  3.30^{+  0.51}_{ -0.45}     $ & $  0.91^{+  0.18}_{ -0.15}     $ & $  2.60^{+  0.71}_{ -0.57}     $ & $  2.60^{+  1.10}_{ -0.78}     $ &$10^{ -8.0}          $ &   5.40 &  -0.3 & fsh  \\
c131  & J141813.3+522414 & $  18.6^{+   6.1}_{  -5.0}     $ & $  12.8^{+   5.1}_{  -4.0}     $ & $<  8.60                       $ & $<  7.27                       $ & $  1.50^{+  0.37}_{ -0.30}     $ & $  0.41^{+  0.13}_{ -0.10}     $ & $<  1.50                       $ & $<  2.50                       $ &$10^{ -8.0}          $ &   6.20 &  -1.0 & fs   \\
c132  & J141813.9+522624 & $  17.5^{+   5.8}_{  -4.7}     $ & $   3.6^{+   3.6}_{  -2.4}     $ & $  13.1^{+   5.1}_{  -4.0}     $ & $<  6.12                       $ & $  1.10^{+  0.30}_{ -0.24}     $ & $  0.09^{+  0.06}_{ -0.04}     $ & $  1.90^{+  0.61}_{ -0.47}     $ & $<  1.70                       $ &$10^{ -8.0}          $ &   5.10 &   0.6 & fh   \\
c133  & J141814.2+522810 & $  24.0^{+   6.4}_{  -5.3}     $ & $  24.1^{+   6.2}_{  -5.1}     $ & $   5.3^{+   4.0}_{  -2.8}     $ & $<  6.19                       $ & $  1.60^{+  0.35}_{ -0.29}     $ & $  0.62^{+  0.15}_{ -0.12}     $ & $  0.78^{+  0.39}_{ -0.27}     $ & $<  1.80                       $ &$10^{ -8.0}          $ &   4.80 &  -0.6 & fs   \\
c134  & J141815.3+523247 & $  43.3^{+   8.3}_{  -7.2}     $ & $  31.3^{+   7.0}_{  -5.9}     $ & $  11.8^{+   5.3}_{  -4.2}     $ & $<  7.96                       $ & $  2.90^{+  0.47}_{ -0.40}     $ & $  0.83^{+  0.17}_{ -0.14}     $ & $  1.80^{+  0.52}_{ -0.41}     $ & $<  2.30                       $ &$10^{ -8.0}          $ &   6.60 &  -0.4 & fs   \\
c135  & J141816.2+522940 & $ 369.2^{+  20.4}_{ -19.3}     $ & $ 287.0^{+  18.0}_{ -17.0}     $ & $  86.6^{+  10.5}_{  -9.5}     $ & $  19.7^{+   5.8}_{  -4.7}     $ & $ 25.00^{+  1.30}_{ -1.30}     $ & $  7.50^{+  0.47}_{ -0.44}     $ & $ 13.00^{+  1.50}_{ -1.40}     $ & $  6.00^{+  1.60}_{ -1.30}     $ &$10^{ -8.0}          $ &   5.20 &  -0.5 & fshu \\
c136  & J141816.3+522524 & $  59.5^{+   9.2}_{  -8.1}     $ & $  41.7^{+   7.8}_{  -6.7}     $ & $  19.5^{+   6.0}_{  -4.9}     $ & $<  7.17                       $ & $  4.00^{+  0.55}_{ -0.49}     $ & $  1.10^{+  0.19}_{ -0.16}     $ & $  2.90^{+  0.72}_{ -0.59}     $ & $<  2.00                       $ &$10^{ -8.0}          $ &   5.90 &  -0.3 & fsh  \\
c137  & J141816.4+523329 & $  26.6^{+   7.2}_{  -6.1}     $ & $  14.7^{+   5.6}_{  -4.4}     $ & $  11.1^{+   5.4}_{  -4.3}     $ & $<  8.91                       $ & $  1.80^{+  0.35}_{ -0.29}     $ & $  0.40^{+  0.11}_{ -0.09}     $ & $  1.70^{+  0.48}_{ -0.38}     $ & $<  2.60                       $ &$10^{ -8.0}          $ &   7.20 &  -0.1 & fs   \\
c138  & J141816.7+522307 & $ 139.6^{+  13.4}_{ -12.4}     $ & $ 104.8^{+  11.6}_{ -10.5}     $ & $  38.8^{+   8.0}_{  -6.9}     $ & $<  9.45                       $ & $  9.60^{+  0.84}_{ -0.77}     $ & $  2.80^{+  0.29}_{ -0.27}     $ & $  6.00^{+  1.00}_{ -0.86}     $ & $<  2.80                       $ &$10^{ -8.0}          $ &   7.40 &  -0.4 & fsh  \\
c139  & J141818.0+523201 & $  28.8^{+   7.1}_{  -6.1}     $ & $  17.3^{+   5.7}_{  -4.5}     $ & $  11.4^{+   5.2}_{  -4.1}     $ & $<  7.49                       $ & $  2.00^{+  0.39}_{ -0.33}     $ & $  0.47^{+  0.13}_{ -0.10}     $ & $  1.80^{+  0.54}_{ -0.42}     $ & $<  2.30                       $ &$10^{ -8.0}          $ &   6.40 &  -0.2 & fs   \\
c140  & J141819.9+522115 & $  25.0^{+   8.0}_{  -6.9}     $ & $   8.6^{+   5.6}_{  -4.4}     $ & $  15.1^{+   6.6}_{  -5.4}     $ & $< 10.51                       $ & $  1.90^{+  0.32}_{ -0.27}     $ & $  0.26^{+  0.07}_{ -0.06}     $ & $  2.60^{+  0.57}_{ -0.48}     $ & $<  3.70                       $ &$10^{ -6.6}          $ &   9.10 &   0.3 & f    \\
c141  & J141820.3+523351 & $  28.9^{+   7.7}_{  -6.6}     $ & $  20.7^{+   6.4}_{  -5.3}     $ & $< 11.96                       $ & $<  9.96                       $ & $  2.00^{+  0.35}_{ -0.30}     $ & $  0.56^{+  0.13}_{ -0.11}     $ & $<  1.80                       $ & $<  3.00                       $ &$10^{ -6.9}          $ &   7.90 &  -1.0 & fs   \\
c142  & J141821.3+522655 & $  21.9^{+   6.4}_{  -5.3}     $ & $  14.4^{+   5.3}_{  -4.2}     $ & $<  9.11                       $ & $<  7.87                       $ & $  1.50^{+  0.34}_{ -0.28}     $ & $  0.40^{+  0.12}_{ -0.09}     $ & $<  1.40                       $ & $<  2.40                       $ &$10^{ -8.0}          $ &   6.00 &  -1.0 & fs   \\
c143  & J141821.3+523254 & $  26.3^{+   7.3}_{  -6.2}     $ & $  14.9^{+   5.7}_{  -4.5}     $ & $  14.2^{+   5.9}_{  -4.8}     $ & $<  4.57                       $ & $  1.80^{+  0.34}_{ -0.29}     $ & $  0.40^{+  0.11}_{ -0.09}     $ & $  2.20^{+  0.55}_{ -0.45}     $ & $<  0.76                       $ &$10^{ -8.0}          $ &   7.40 &   0.0 & fsh  \\
c144  & J141821.7+522955 & $  20.6^{+   6.4}_{  -5.3}     $ & $  21.5^{+   6.1}_{  -5.0}     $ & $<  9.04                       $ & $<  8.35                       $ & $  1.40^{+  0.31}_{ -0.26}     $ & $  0.58^{+  0.14}_{ -0.11}     $ & $<  1.30                       $ & $<  2.50                       $ &$10^{ -8.0}          $ &   6.10 &  -1.0 & fs   \\
c145  & J141822.0+522650 & $  36.2^{+   7.7}_{  -6.6}     $ & $  25.2^{+   6.4}_{  -5.3}     $ & $  10.8^{+   5.1}_{  -4.0}     $ & $   7.7^{+   4.4}_{  -3.3}     $ & $  2.50^{+  0.44}_{ -0.38}     $ & $  0.69^{+  0.15}_{ -0.13}     $ & $  1.70^{+  0.54}_{ -0.42}     $ & $  2.40^{+  0.98}_{ -0.72}     $ &$10^{ -8.0}          $ &   6.10 &  -0.4 & fs   \\
c146  & J141822.4+523607 & $ 163.6^{+  15.1}_{ -14.0}     $ & $ 110.7^{+  12.3}_{ -11.2}     $ & $  58.5^{+  10.0}_{  -8.9}     $ & $  18.9^{+   6.6}_{  -5.5}     $ & $ 12.00^{+  0.90}_{ -0.84}     $ & $  3.10^{+  0.30}_{ -0.28}     $ & $  9.50^{+  1.20}_{ -1.10}     $ & $  6.50^{+  1.40}_{ -1.20}     $ &$10^{ -8.0}          $ &   9.80 &  -0.3 & fshu \\
c147  & J141822.8+522710 & $  22.1^{+   6.6}_{  -5.4}     $ & $  22.0^{+   6.2}_{  -5.1}     $ & $<  8.82                       $ & $<  7.58                       $ & $  1.60^{+  0.34}_{ -0.28}     $ & $  0.61^{+  0.14}_{ -0.12}     $ & $<  1.30                       $ & $<  2.30                       $ &$10^{ -8.0}          $ &   6.20 &  -1.0 & fs   \\
c148  & J141823.0+522114 & $  43.1^{+   9.4}_{  -8.3}     $ & $  16.5^{+   6.4}_{  -5.3}     $ & $  25.5^{+   7.7}_{  -6.6}     $ & $< 11.93                       $ & $  3.20^{+  0.43}_{ -0.39}     $ & $  0.48^{+  0.11}_{ -0.09}     $ & $  4.30^{+  0.75}_{ -0.65}     $ & $<  4.10                       $ &$10^{ -8.0}          $ &   9.40 &   0.2 & fh   \\
c149  & J141824.9+522330 & $ 290.0^{+  18.5}_{ -17.5}     $ & $ 156.3^{+  13.8}_{ -12.8}     $ & $ 138.1^{+  13.2}_{ -12.2}     $ & $  33.6^{+   7.4}_{  -6.3}     $ & $ 21.00^{+  1.30}_{ -1.20}     $ & $  4.50^{+  0.38}_{ -0.35}     $ & $ 23.00^{+  2.00}_{ -1.90}     $ & $ 12.00^{+  2.10}_{ -1.80}     $ &$10^{ -8.0}          $ &   8.10 &   0.0 & fshu \\
c150  & J141825.5+522349 & $ 193.6^{+  15.5}_{ -14.4}     $ & $ 140.7^{+  13.2}_{ -12.1}     $ & $  55.8^{+   9.2}_{  -8.1}     $ & $  19.8^{+   6.2}_{  -5.1}     $ & $ 14.00^{+  1.00}_{ -0.97}     $ & $  4.00^{+  0.36}_{ -0.33}     $ & $  9.10^{+  1.30}_{ -1.10}     $ & $  6.80^{+  1.60}_{ -1.30}     $ &$10^{ -8.0}          $ &   7.90 &  -0.4 & fshu \\
c151  & J141826.3+522818 & $  98.1^{+  11.5}_{ -10.4}     $ & $  89.6^{+  10.8}_{  -9.7}     $ & $  22.6^{+   6.6}_{  -5.4}     $ & $<  9.18                       $ & $  6.60^{+  0.70}_{ -0.64}     $ & $  2.40^{+  0.27}_{ -0.24}     $ & $  3.50^{+  0.77}_{ -0.64}     $ & $<  2.80                       $ &$10^{ -8.0}          $ &   6.60 &  -0.6 & fsh  \\
c152  & J141826.4+523235 & $  17.7^{+   6.9}_{  -5.8}     $ & $  19.8^{+   6.3}_{  -5.2}     $ & $< 12.67                       $ & $< 10.21                       $ & $  1.20^{+  0.25}_{ -0.21}     $ & $  0.54^{+  0.13}_{ -0.10}     $ & $<  1.90                       $ & $<  3.10                       $ &$10^{ -8.0}          $ &   7.80 &  -1.0 & sf   \\
c153  & J141826.5+522559 & $  31.2^{+   7.5}_{  -6.4}     $ & $  11.7^{+   5.2}_{  -4.1}     $ & $  19.7^{+   6.3}_{  -5.2}     $ & $<  8.55                       $ & $  2.20^{+  0.40}_{ -0.34}     $ & $  0.33^{+  0.10}_{ -0.08}     $ & $  3.20^{+  0.73}_{ -0.60}     $ & $<  2.70                       $ &$10^{ -8.0}          $ &   7.00 &   0.3 & fsh  \\
c154  & J141829.7+522709 & $  27.2^{+   7.4}_{  -6.3}     $ & $  10.8^{+   5.2}_{  -4.1}     $ & $  18.4^{+   6.3}_{  -5.2}     $ & $   8.7^{+   4.8}_{  -3.7}     $ & $  2.00^{+  0.36}_{ -0.31}     $ & $  0.30^{+  0.09}_{ -0.07}     $ & $  3.00^{+  0.69}_{ -0.57}     $ & $  2.90^{+  1.00}_{ -0.76}     $ &$10^{ -8.0}          $ &   7.20 &   0.3 & fh   \\
c155  & J141830.2+522212 & $1118.2^{+  34.9}_{ -33.9}     $ & $ 885.5^{+  31.0}_{ -30.0}     $ & $ 265.4^{+  17.9}_{ -16.9}     $ & $  68.5^{+  10.0}_{  -8.9}     $ & $ 81.00^{+  2.40}_{ -2.40}     $ & $ 25.00^{+  0.86}_{ -0.83}     $ & $ 44.00^{+  2.70}_{ -2.60}     $ & $ 24.00^{+  3.00}_{ -2.70}     $ &$10^{ -8.0}          $ &   9.50 &  -0.5 & fshu \\
c156  & J141832.8+522349 & $  83.4^{+  11.4}_{ -10.3}     $ & $  61.9^{+   9.6}_{  -8.6}     $ & $  23.8^{+   7.3}_{  -6.2}     $ & $  10.5^{+   5.4}_{  -4.3}     $ & $  6.40^{+  0.68}_{ -0.62}     $ & $  1.90^{+  0.24}_{ -0.22}     $ & $  4.10^{+  0.78}_{ -0.66}     $ & $  3.80^{+  1.10}_{ -0.88}     $ &$10^{ -8.0}          $ &   8.90 &  -0.4 & fsh  \\
c157  & J141837.9+522034 & $  25.8^{+   9.8}_{  -8.8}     $ & $  19.2^{+   7.4}_{  -6.3}     $ & $<  8.55                       $ & $< 14.74                       $ & $  2.40^{+  0.31}_{ -0.27}     $ & $  0.72^{+  0.13}_{ -0.11}     $ & $<  1.70                       $ & $<  6.70                       $ &$10^{ -5.6}          $ &  11.50 &  -1.0 & s    \\
c158  & J141838.1+522358 & $ 129.2^{+  12.8}_{ -11.7}     $ & $  99.3^{+  11.2}_{ -10.2}     $ & $  32.6^{+   7.2}_{  -6.1}     $ & $<  7.05                       $ & $ 39.00^{+  3.60}_{ -3.30}     $ & $ 11.00^{+  1.20}_{ -1.10}     $ & $ 22.00^{+  4.20}_{ -3.60}     $ & $<  9.70                       $ &$10^{ -8.0}          $ &   9.50 &  -0.5 & fsh  \\
\hline
\end{tabular}
\end{center}
\end{table}
\end{landscape}

\appendix
\section{Source detection in HDF-N and ELAIS-N1 fields}

As described above, in order to make a fair comparison between the number counts in the GWS with other fields, we have applied our source detection technique to two comparison fields, ELAIS-N1 (Manners et al. 2003) and the 2Ms HDF-N survey (Alexander et al. 2003). For completeness, we describe the results of our analysis in more detail here, and compare them to the previous work. 

\subsection{ELAIS-N1}

As stated above, applying our analysis method to the ELAIS-N1 data resulted in 145 band-merged sources, with 132 FB, 116, SB, 88 HB and 29 UB. For comparison, in the published catalogue of M03 there are 124 band-merged ACIS-I sources (i.e. excluding 6 ACIS-S sources identified by M03; we do not consider the S-chips in our analysis). Matching the two source lists show that we identify all the M03 sources, but detect 21 additional sources. These extra sources are shown in Table~\ref{tab:en1}. They were visually inspected to ensure that they were likely to be real and not artifacts. 

There are a number of differences in the analysis presented by M03 and ours. M03 used the pipeline-processed evt2 files, whereas we used our own processing. We have filtered out background flares, so our exposure time of 70.0ks is shorter than the 71.2ks of M03. They performed source detection on the binned image, whereas we use the raw one. Finally, we adopted an upper limit of 7 keV whereas their source detection used the data up to 8 keV. Given the fact that we detect all the ACIS-I sources in M03 and 21 additional sources we are confident that our procedure has resulted in a more comprehensive point source catalogue in the case
of ELAIS-N1. 

\subsection{HDF-N}

Running our source procedure at a threshold of $4 \times 10^{-6}$ on the 2 Ms exposure of the HDF-N resulted in  536 band-merged sources. There are  485 FB, 432 SB, 312 HB and 
187 UB sources. Matching this to the main catalogue of Alexander et al. (2003), which contains
503 sources, gives 461 sources in common. We detect 75 sources that A03 do not, and these are listed in Table~\ref{tab:hdfn}. A03 detect 42 sources which are not significant in our analysis.  We note that our analysis does not necessarily cast doubt on the reality of these 42 sources - weak and slightly extended sources could be missed by our method, and we use circular aperture extraction rather than the true PSF shape. It is highly likely, however, that the additional 75 sources we detect that A03 do not are also real. A large fraction ($\sim 40$~per cent) of the non-common sources are close to the chip edges or the chip gaps, where any detection procedure will have difficulty and produce mixed results. 

Again there are differences between out basic analysis and that of A03 which can also account for the differences in the source lists. Our screening was slightly more stringent (1.86 Ms exposure versus 1.95 Ms in A03), and A03 used an upper bound of 8 keV, rather than 7 keV. They also performed source detection in a total of 7 bands, not 4, and repeated the analysis for the restricted ACIS grade set, as well as the standard ASCA grade set we use here. 

\begin{landscape}
\begin{table}
\centering
\caption{Additional sources detected in ELAIS-N1.
Col.(1): Source catalogue number;
Col.(2): Chandra object name;
Col.(3): Full band (0.5-7 keV) counts;
Col.(4): Soft band (0.5-2 keV) counts;
Col.(5): Hard band (2-7 keV) counts;
Col.(6): Ultra-hard band (4-7 keV) counts;
Col.(7): 0.5-10 keV flux (all fluxes $10^{-15}$ erg cm$^{-2}$ s$^{-1}$);
Col.(8): 0.5-2 keV flux;
Col.(9): 2-10 keV flux;
Col.(10): 5-10 keV flux;
Col.(11): $p_{\rm min}$ is the lowest false detection probability found for 
the four bands. A probability
of $10^{-8}$ is assigned if the Poisson probability is less than this value. 
Col.(12); Hardness ratio = (H-S)/(H+S) where H and S are the 2-7 keV and 0.5- 2 keV counts, corrected to on-axis values. 
Col.(13); Off-axis angle in arcmin. 
Col.(14); Flags fshu=source detected at
$<4 \times 10^{-6}$ probability in this band, where the bands are
f=full, s=soft, h=hard, u=ultrahard. The first band quoted is the one with the lowest
probability (i.e. that quoted in column 11). 
\label{tab:en1}}
\begin{center}
\begin{tabular}{@{}ccrrrrrrrrcccc@{}}
\hline
Cat & CXO EN1 & FB & SB & HB & UB & 0.5-10.0 & 0.5-2.0 & 2-10 & 5-10 & 
  $p_{\rm min}$ & OAA & HR& Flags \\
No. & (2000) & cts & cts & cts & cts & flux & flux & flux & flux & & ($\prime$) & &  \\
(1) & (2) & (3) & (4) & (5) & (6) & (7) & (8) & (9) & (10) & (11) & (12) & (13) & (14)\\
\hline
a6    & J160929.1+542938 & $  12.9^{+   5.3}_{  -4.2}     $ & $   5.5^{+   4.0}_{  -2.8}     $ & $   6.7^{+   4.3}_{  -3.1}     $ & $<  6.16                       $ & $  3.60^{+  1.10}_{ -0.83}     $ & $  0.40^{+  0.20}_{ -0.14}     $ & $  2.90^{+  1.20}_{ -0.90}     $ & $<  5.30                       $ &$10^{ -6.1}          $ &   8.30 &   0.1 & f    \\
a21   & J160946.6+543008 & $   8.4^{+   4.3}_{  -3.1}     $ & $   5.2^{+   3.6}_{  -2.4}     $ & $<  5.00                       $ & $<  4.30                       $ & $  2.10^{+  0.91}_{ -0.66}     $ & $  0.35^{+  0.21}_{ -0.14}     $ & $<  1.90                       $ & $<  3.20                       $ &$10^{ -6.7}          $ &   5.80 &  -1.0 & sf   \\
a39   & J161001.9+544148 & $   7.9^{+   4.8}_{  -3.7}     $ & $   8.1^{+   4.4}_{  -3.3}     $ & $<  7.97                       $ & $<  7.29                       $ & $  2.10^{+  0.72}_{ -0.54}     $ & $  0.56^{+  0.23}_{ -0.17}     $ & $<  3.10                       $ & $<  6.10                       $ &$10^{ -5.6}          $ &   8.90 &  -1.0 & fs   \\
a51   & J161010.7+543556 & $   5.6^{+   3.6}_{  -2.4}     $ & $<  2.82                       $ & $   4.7^{+   3.4}_{  -2.2}     $ & $<  3.75                       $ & $  1.30^{+  0.80}_{ -0.53}     $ & $<  0.18                       $ & $  1.70^{+  1.20}_{ -0.74}     $ & $<  2.50                       $ &$10^{ -8.0}          $ &   2.90 &   1.0 & fh   \\
a55   & J161014.4+543359 & $   2.8^{+   2.9}_{  -1.6}     $ & $<  2.90                       $ & $   2.8^{+   2.9}_{  -1.6}     $ & $<  2.86                       $ & $  0.69^{+  0.67}_{ -0.37}     $ & $<  0.19                       $ & $  1.10^{+  1.10}_{ -0.58}     $ & $<  2.00                       $ &$10^{ -5.5}          $ &   1.00 &   1.0 & h    \\
a61   & J161017.5+543747 & $   4.2^{+   3.4}_{  -2.2}     $ & $   3.7^{+   3.2}_{  -1.9}     $ & $<  4.46                       $ & $<  3.62                       $ & $  1.00^{+  0.69}_{ -0.44}     $ & $  0.23^{+  0.19}_{ -0.11}     $ & $<  1.60                       $ & $<  2.50                       $ &$10^{ -5.8}          $ &   4.40 &  -1.0 & s    \\
a62   & J161018.2+543232 & $   3.7^{+   3.2}_{  -1.9}     $ & $   3.9^{+   3.2}_{  -1.9}     $ & $<  2.85                       $ & $<  2.83                       $ & $  0.87^{+  0.69}_{ -0.42}     $ & $  0.24^{+  0.19}_{ -0.11}     $ & $<  0.99                       $ & $<  1.90                       $ &$10^{ -5.7}          $ &   0.90 &  -1.0 & s    \\
a69   & J161020.7+544144 & $  12.8^{+   5.3}_{  -4.2}     $ & $  10.8^{+   4.7}_{  -3.5}     $ & $<  8.33                       $ & $<  6.57                       $ & $  3.30^{+  0.99}_{ -0.78}     $ & $  0.74^{+  0.27}_{ -0.20}     $ & $<  3.20                       $ & $<  5.40                       $ &$10^{ -6.1}          $ &   8.40 &  -1.0 & s    \\
a76   & J161026.0+543021 & $   7.6^{+   4.0}_{  -2.8}     $ & $   5.7^{+   3.6}_{  -2.4}     $ & $<  3.69                       $ & $<  3.78                       $ & $  1.80^{+  0.91}_{ -0.63}     $ & $  0.37^{+  0.22}_{ -0.14}     $ & $<  1.30                       $ & $<  2.60                       $ &$10^{ -8.0}          $ &   3.10 &  -1.0 & fs   \\
a77   & J161026.1+543253 & $   5.8^{+   3.6}_{  -2.4}     $ & $   4.9^{+   3.4}_{  -2.2}     $ & $<  2.84                       $ & $<  2.86                       $ & $  1.50^{+  0.92}_{ -0.61}     $ & $  0.34^{+  0.23}_{ -0.15}     $ & $<  1.10                       $ & $<  2.30                       $ &$10^{ -6.9}          $ &   1.00 &  -1.0 & f    \\
a78   & J161026.1+542528 & $   7.5^{+   4.4}_{  -3.3}     $ & $   7.1^{+   4.1}_{  -2.9}     $ & $<  6.80                       $ & $<  5.70                       $ & $  2.50^{+  1.00}_{ -0.74}     $ & $  0.61^{+  0.28}_{ -0.20}     $ & $<  3.40                       $ & $<  5.90                       $ &$10^{ -5.7}          $ &   7.90 &  -1.0 & s    \\
a82   & J161029.9+543315 & $   2.8^{+   2.9}_{  -1.6}     $ & $   2.9^{+   2.9}_{  -1.6}     $ & $<  2.82                       $ & $<  2.85                       $ & $  0.66^{+  0.65}_{ -0.36}     $ & $  0.18^{+  0.18}_{ -0.10}     $ & $<  1.00                       $ & $<  2.10                       $ &$10^{ -6.1}          $ &   1.40 &  -1.0 & s    \\
a87   & J161035.4+543558 & $   6.5^{+   3.8}_{  -2.6}     $ & $   4.7^{+   3.4}_{  -2.2}     $ & $<  3.66                       $ & $<  3.73                       $ & $  1.60^{+  0.85}_{ -0.58}     $ & $  0.30^{+  0.21}_{ -0.13}     $ & $<  1.30                       $ & $<  2.70                       $ &$10^{ -8.0}          $ &   3.40 &  -1.0 & fs   \\
a88   & J161037.2+542556 & $  10.6^{+   5.0}_{  -3.8}     $ & $   5.7^{+   4.0}_{  -2.8}     $ & $<  7.08                       $ & $<  6.19                       $ & $  3.10^{+  1.00}_{ -0.78}     $ & $  0.43^{+  0.21}_{ -0.15}     $ & $<  3.10                       $ & $<  5.60                       $ &$10^{ -5.6}          $ &   7.80 &  -1.0 & fs   \\
a100  & J161044.9+543003 & $   5.7^{+   3.8}_{  -2.6}     $ & $   2.3^{+   2.9}_{  -1.6}     $ & $   3.2^{+   3.2}_{  -1.9}     $ & $<  4.40                       $ & $  1.40^{+  0.77}_{ -0.53}     $ & $  0.15^{+  0.15}_{ -0.08}     $ & $  1.20^{+  0.99}_{ -0.59}     $ & $<  3.30                       $ &$10^{ -6.1}          $ &   4.90 &   0.2 & f    \\
a103  & J161045.9+543046 & $   5.7^{+   3.8}_{  -2.6}     $ & $   3.3^{+   3.2}_{  -1.9}     $ & $<  5.21                       $ & $<  4.46                       $ & $  1.40^{+  0.77}_{ -0.53}     $ & $  0.22^{+  0.17}_{ -0.10}     $ & $<  2.00                       $ & $<  3.40                       $ &$10^{ -6.1}          $ &   4.50 &  -1.0 & f    \\
a114  & J161048.8+543458 & $   5.1^{+   3.6}_{  -2.4}     $ & $<  3.58                       $ & $   4.3^{+   3.4}_{  -2.2}     $ & $<  3.59                       $ & $  1.30^{+  0.76}_{ -0.50}     $ & $<  0.23                       $ & $  1.70^{+  1.20}_{ -0.73}     $ & $<  2.70                       $ &$10^{ -6.2}          $ &   4.50 &   1.0 & fh   \\
a125  & J161056.3+543208 & $   4.5^{+   3.6}_{  -2.4}     $ & $   4.1^{+   3.4}_{  -2.2}     $ & $<  5.05                       $ & $<  4.30                       $ & $  1.10^{+  0.68}_{ -0.45}     $ & $  0.28^{+  0.19}_{ -0.12}     $ & $<  1.90                       $ & $<  3.40                       $ &$10^{ -5.5}          $ &   5.40 &  -1.0 & s    \\
a137  & J161111.6+543424 & $  15.4^{+   5.6}_{  -4.4}     $ & $   6.0^{+   4.0}_{  -2.8}     $ & $<  7.70                       $ & $<  6.03                       $ & $  4.10^{+  1.10}_{ -0.90}     $ & $  0.42^{+  0.21}_{ -0.14}     $ & $<  3.00                       $ & $<  4.90                       $ &$10^{ -5.8}          $ &   7.50 &  -1.0 & f    \\
a140  & J161116.5+543235 & $  14.0^{+   5.7}_{  -4.5}     $ & $<  7.67                       $ & $  10.5^{+   5.0}_{  -3.8}     $ & $   6.3^{+   4.1}_{  -2.9}     $ & $  3.90^{+  1.10}_{ -0.85}     $ & $<  0.57                       $ & $  4.60^{+  1.50}_{ -1.20}     $ & $  5.90^{+  2.70}_{ -1.90}     $ &$10^{ -6.6}          $ &   8.20 &   1.0 & hf   \\
a142  & J161120.6+543313 & $   8.4^{+   4.4}_{  -3.3}     $ & $   5.7^{+   3.8}_{  -2.6}     $ & $<  6.32                       $ & $<  5.74                       $ & $  7.70^{+  3.10}_{ -2.30}     $ & $  1.40^{+  0.74}_{ -0.51}     $ & $<  8.70                       $ & $< 17.00                       $ &$10^{ -5.6}          $ &   8.80 &  -1.0 & s    \\
\hline
\end{tabular}
\end{center}
\end{table}
\end{landscape}

\begin{landscape}
\begin{table}
\centering
\caption{Additional sources detected in HDF-N (see text).
Col.(1): Source catalogue number;
Col.(2): Chandra object name;
Col.(3): Full band (0.5-7 keV) counts;
Col.(4): Soft band (0.5-2 keV) counts;
Col.(5): Hard band (2-7 keV) counts;
Col.(6): Ultra-hard band (4-7 keV) counts;
Col.(7): 0.5-10 keV flux (all fluxes $10^{-15}$ erg cm$^{-2}$ s$^{-1}$);
Col.(8): 0.5-2 keV flux;
Col.(9): 2-10 keV flux;
Col.(10): 5-10 keV flux;
Col.(11): $p_{\rm min}$ is the lowest false detection probability found for 
the four bands. A probability
of $10^{-8}$ is assigned if the Poisson probability is less than this value. 
Col.(12); Off-axis angle in arcmin. 
Col.(13); Hardness ratio = (H-S)/(H+S) where H and S are the 2-7 keV and 0.5- 2 keV counts, corrected to on-axis values. 
Col.(14); Flags fshu=source detected at
$<4 \times 10^{-6}$ probability in this band, where the bands are
f=full, s=soft, h=hard, u=ultrahard. The first band quoted is the one with the lowest
probability (i.e. that quoted in column 11). 
\label{tab:hdfn}}
\begin{center}
\begin{tabular}{@{}ccrrrrrrrrcccc@{}}
\hline
Cat & CXO HDFN & FB & SB & HB & UB & 0.5-10.0 & 0.5-2.0 & 2-10 & 5-10 & 
  $p_{\rm min}$ & OAA & HR & Flags \\
No. & (2000) & cts & cts & cts & cts & flux & flux & flux & flux & & ($\prime$) & &  \\
(1) & (2) & (3) & (4) & (5) & (6) & (7) & (8) & (9) & (10) & (11) & (12) & (13) & (14)\\
\hline
f9    & J123523.2+621302 & $  65.5^{+  11.9}_{ -10.9}     $ & $  40.8^{+   9.1}_{  -8.0}     $ & $<  8.86                       $ & $< 15.51                       $ & $  6.50^{+  0.66}_{ -0.60}     $ & $  1.10^{+  0.15}_{ -0.13}     $ & $<  0.56                       $ & $<  4.90                       $ &$10^{ -8.0}          $ &   9.90 &  -0.2 & fs   \\
f10   & J123524.0+621359 & $  43.8^{+  12.3}_{ -11.2}     $ & $  22.5^{+   8.8}_{  -7.7}     $ & $  27.1^{+  10.2}_{  -9.1}     $ & $< 15.51                       $ & $  2.10^{+  0.20}_{ -0.19}     $ & $  0.29^{+  0.04}_{ -0.04}     $ & $  2.00^{+  0.24}_{ -0.22}     $ & $<  2.30                       $ &$10^{ -8.0}          $ &   9.70 &   0.1 & fh   \\
f14   & J123529.3+621223 & $  35.5^{+   9.7}_{  -8.6}     $ & $  14.8^{+   6.9}_{  -5.8}     $ & $  18.0^{+   7.8}_{  -6.7}     $ & $< 13.55                       $ & $  3.60^{+  0.46}_{ -0.41}     $ & $  0.39^{+  0.08}_{ -0.07}     $ & $  2.80^{+  0.48}_{ -0.42}     $ & $<  4.30                       $ &$10^{ -8.0}          $ &   9.30 &   0.1 & fs   \\
f16   & J123532.0+621458 & $  32.1^{+  12.7}_{ -11.7}     $ & $< 13.55                       $ & $  34.3^{+  11.4}_{ -10.4}     $ & $< 13.55                       $ & $  0.60^{+  0.06}_{ -0.05}     $ & $<  0.07                       $ & $  1.00^{+  0.11}_{ -0.10}     $ & $<  0.79                       $ &$10^{ -5.5}          $ &   8.80 &   1.0 & h    \\
f17   & J123532.4+621120 & $  19.5^{+   8.3}_{  -7.2}     $ & $  19.9^{+   6.9}_{  -5.8}     $ & $< 16.18                       $ & $< 13.14                       $ & $  2.60^{+  0.41}_{ -0.35}     $ & $  0.70^{+  0.14}_{ -0.12}     $ & $<  3.10                       $ & $<  5.40                       $ &$10^{ -5.5}          $ &   9.20 &  -1.0 & s    \\
f41   & J123551.8+620939 & $  22.7^{+   7.6}_{  -6.5}     $ & $< 10.99                       $ & $  18.7^{+   6.8}_{  -5.7}     $ & $  10.4^{+   5.4}_{  -4.3}     $ & $  2.40^{+  0.43}_{ -0.37}     $ & $<  0.31                       $ & $  3.10^{+  0.63}_{ -0.53}     $ & $  3.50^{+  1.00}_{ -0.80}     $ &$10^{ -5.7}          $ &   7.90 &   1.0 & hf   \\
f42   & J123552.5+622227 & $  21.0^{+   7.8}_{  -6.8}     $ & $  18.6^{+   6.4}_{  -5.3}     $ & $< 15.02                       $ & $< 11.24                       $ & $  2.90^{+  0.49}_{ -0.42}     $ & $  0.68^{+  0.15}_{ -0.12}     $ & $<  3.00                       $ & $<  4.70                       $ &$10^{ -5.9}          $ &  10.50 &  -1.0 & s    \\
f49   & J123555.1+620843 & $  30.9^{+   8.4}_{  -7.3}     $ & $< 11.46                       $ & $  23.6^{+   7.4}_{  -6.3}     $ & $< 11.91                       $ & $  3.40^{+  0.53}_{ -0.46}     $ & $<  0.34                       $ & $  4.00^{+  0.74}_{ -0.63}     $ & $<  4.00                       $ &$10^{ -6.7}          $ &   8.20 &   1.0 & f    \\
f50   & J123555.3+620835 & $  23.7^{+   8.1}_{  -7.0}     $ & $< 11.66                       $ & $  24.1^{+   7.5}_{  -6.4}     $ & $< 11.91                       $ & $  2.60^{+  0.42}_{ -0.36}     $ & $<  0.34                       $ & $  4.00^{+  0.72}_{ -0.61}     $ & $<  3.90                       $ &$10^{ -8.0}          $ &   8.20 &   1.0 & hf   \\
f60   & J123558.0+621354 & $  31.1^{+  10.4}_{  -9.4}     $ & $  17.1^{+   7.8}_{  -6.8}     $ & $  19.2^{+   8.9}_{  -7.8}     $ & $  19.5^{+   7.9}_{  -6.8}     $ & $  0.40^{+  0.05}_{ -0.04}     $ & $  0.06^{+  0.01}_{ -0.01}     $ & $  0.38^{+  0.05}_{ -0.05}     $ & $  0.80^{+  0.14}_{ -0.12}     $ &$10^{ -8.0}          $ &   5.80 &   0.1 & fs   \\
f66   & J123600.5+621215 & $  25.4^{+   9.9}_{  -8.8}     $ & $< 17.68                       $ & $  35.4^{+   9.7}_{  -8.6}     $ & $  28.1^{+   8.4}_{  -7.3}     $ & $  0.32^{+  0.04}_{ -0.04}     $ & $<  0.06                       $ & $  0.70^{+  0.09}_{ -0.08}     $ & $  1.10^{+  0.18}_{ -0.16}     $ &$10^{ -8.0}          $ &   5.80 &   1.0 & hfu  \\
f74   & J123604.7+622208 & $  22.4^{+  10.6}_{  -9.5}     $ & $  22.5^{+   8.3}_{  -7.2}     $ & $< 17.59                       $ & $< 17.95                       $ & $  0.57^{+  0.07}_{ -0.06}     $ & $  0.15^{+  0.02}_{ -0.02}     $ & $<  0.66                       $ & $<  1.40                       $ &$10^{ -6.5}          $ &   9.40 &  -1.0 & s    \\
f91   & J123608.8+621113 & $  16.0^{+   8.6}_{  -7.5}     $ & $  14.8^{+   7.1}_{  -6.0}     $ & $< 18.50                       $ & $< 15.71                       $ & $  0.22^{+  0.03}_{ -0.03}     $ & $  0.05^{+  0.01}_{ -0.01}     $ & $<  0.37                       $ & $<  0.66                       $ &$10^{ -7.2}          $ &   5.40 &  -1.0 & s    \\
f93   & J123610.2+620748 & $  21.2^{+   8.2}_{  -7.1}     $ & $< 13.27                       $ & $  16.9^{+   7.2}_{  -6.1}     $ & $  11.4^{+   6.0}_{  -4.9}     $ & $  1.30^{+  0.21}_{ -0.18}     $ & $<  0.22                       $ & $  1.60^{+  0.30}_{ -0.25}     $ & $  2.20^{+  0.55}_{ -0.45}     $ &$10^{ -6.2}          $ &   7.70 &   1.0 & h    \\
f107  & J123614.1+622523 & $  65.6^{+  11.8}_{ -10.8}     $ & $< 16.17                       $ & $  54.3^{+  10.4}_{  -9.3}     $ & $  26.1^{+   7.8}_{  -6.7}     $ & $  6.60^{+  0.67}_{ -0.61}     $ & $<  0.43                       $ & $  8.40^{+  1.00}_{ -0.90}     $ & $  8.90^{+  1.50}_{ -1.30}     $ &$10^{ -8.0}          $ &  11.90 &   1.0 & fhu  \\
f112  & J123615.1+621008 & $  30.9^{+   9.6}_{  -8.6}     $ & $   9.2^{+   6.6}_{  -5.4}     $ & $  17.6^{+   8.1}_{  -7.0}     $ & $< 15.39                       $ & $  0.38^{+  0.05}_{ -0.04}     $ & $  0.03^{+  0.01}_{ -0.01}     $ & $  0.34^{+  0.06}_{ -0.05}     $ & $<  0.58                       $ &$10^{ -7.2}          $ &   5.50 &   0.3 & f    \\
f121  & J123618.3+621550 & $  29.2^{+   8.8}_{  -7.7}     $ & $  19.5^{+   7.1}_{  -6.1}     $ & $  10.8^{+   6.7}_{  -5.6}     $ & $< 13.62                       $ & $  0.35^{+  0.05}_{ -0.04}     $ & $  0.06^{+  0.01}_{ -0.01}     $ & $  0.20^{+  0.04}_{ -0.04}     $ & $<  0.49                       $ &$10^{ -8.0}          $ &   3.80 &  -0.3 & f    \\
f128  & J123620.3+622507 & $  20.3^{+  12.7}_{ -11.6}     $ & $  23.8^{+   9.5}_{  -8.5}     $ & $< 10.54                       $ & $< 10.54                       $ & $  0.75^{+  0.07}_{ -0.06}     $ & $  0.23^{+  0.03}_{ -0.03}     $ & $<  0.57                       $ & $<  1.20                       $ &$10^{ -5.7}          $ &  11.40 &  -1.0 & s    \\
f136  & J123621.8+621323 & $  12.9^{+   6.2}_{  -5.1}     $ & $< 10.28                       $ & $  10.1^{+   5.6}_{  -4.4}     $ & $   5.8^{+   4.7}_{  -3.5}     $ & $  0.23^{+  0.05}_{ -0.04}     $ & $<  0.05                       $ & $  0.28^{+  0.08}_{ -0.06}     $ & $  0.32^{+  0.12}_{ -0.09}     $ &$10^{ -8.0}          $ &   3.10 &   1.0 & fh   \\
f139  & J123622.1+620517 & $  16.4^{+   7.7}_{  -6.6}     $ & $  13.6^{+   6.1}_{  -5.0}     $ & $< 15.25                       $ & $< 12.33                       $ & $  1.90^{+  0.32}_{ -0.28}     $ & $  0.41^{+  0.10}_{ -0.08}     $ & $<  2.50                       $ & $<  4.30                       $ &$10^{ -6.1}          $ &   9.30 &  -1.0 & s    \\
f159  & J123627.2+622231 & $  36.2^{+  11.5}_{ -10.5}     $ & $<  9.60                       $ & $  32.4^{+  10.1}_{  -9.0}     $ & $  18.1^{+   8.0}_{  -6.9}     $ & $  0.74^{+  0.08}_{ -0.07}     $ & $<  0.05                       $ & $  1.00^{+  0.12}_{ -0.11}     $ & $  1.20^{+  0.19}_{ -0.17}     $ &$10^{ -7.2}          $ &   8.70 &   1.0 & fh   \\
f163  & J123627.8+621448 & $  12.3^{+   6.1}_{  -5.0}     $ & $<  9.82                       $ & $   5.7^{+   5.1}_{  -4.0}     $ & $   1.1^{+   4.1}_{  -2.9}     $ & $  0.15^{+  0.04}_{ -0.03}     $ & $<  0.03                       $ & $  0.11^{+  0.03}_{ -0.03}     $ & $  0.04^{+  0.02}_{ -0.01}     $ &$10^{ -8.0}          $ &   2.40 &   1.0 & fh   \\
f177  & J123630.3+621332 & $< 11.35                       $ & $   6.3^{+   4.4}_{  -3.3}     $ & $< 10.13                       $ & $<  9.70                       $ & $<  0.13                       $ & $  0.02^{+  0.01}_{ -0.01}     $ & $<  0.18                       $ & $<  0.34                       $ &$10^{ -6.1}          $ &   2.10 &  -1.0 & s    \\
f183  & J123632.4+621104 & $  15.6^{+   6.4}_{  -5.3}     $ & $   6.9^{+   4.7}_{  -3.5}     $ & $< 12.20                       $ & $< 10.41                       $ & $  0.19^{+  0.04}_{ -0.04}     $ & $  0.02^{+  0.01}_{ -0.01}     $ & $<  0.22                       $ & $<  0.37                       $ &$10^{ -6.6}          $ &   3.50 &  -1.0 & fs   \\
f188  & J123633.3+621408 & $  14.6^{+   5.9}_{  -4.8}     $ & $  11.4^{+   5.1}_{  -4.0}     $ & $<  4.71                       $ & $<  9.05                       $ & $  0.17^{+  0.04}_{ -0.04}     $ & $  0.04^{+  0.01}_{ -0.01}     $ & $<  0.03                       $ & $<  0.32                       $ &$10^{ -7.2}          $ &   1.70 &  -0.3 & f    \\
f193  & J123633.8+622039 & $  54.8^{+  12.3}_{ -11.2}     $ & $  21.2^{+   8.7}_{  -7.6}     $ & $  39.6^{+  10.6}_{  -9.5}     $ & $< 17.51                       $ & $  0.82^{+  0.08}_{ -0.07}     $ & $  0.09^{+  0.01}_{ -0.01}     $ & $  0.91^{+  0.11}_{ -0.09}     $ & $<  0.78                       $ &$10^{ -8.0}          $ &   6.70 &   0.3 & fh   \\
f206  & J123636.0+622729 & $  37.9^{+  13.2}_{ -12.1}     $ & $  39.5^{+  10.4}_{  -9.4}     $ & $<  9.59                       $ & $<  9.59                       $ & $  2.30^{+  0.21}_{ -0.19}     $ & $  0.65^{+  0.08}_{ -0.07}     $ & $<  0.89                       $ & $<  2.00                       $ &$10^{ -5.7}          $ &  13.40 &  -1.0 & s    \\
f213  & J123637.3+622546 & $  80.8^{+  18.7}_{ -17.6}     $ & $< 16.26                       $ & $  59.5^{+  15.7}_{ -14.7}     $ & $< 16.26                       $ & $  1.80^{+  0.11}_{ -0.10}     $ & $<  0.09                       $ & $  2.00^{+  0.15}_{ -0.14}     $ & $<  1.10                       $ &$10^{ -6.9}          $ &  11.70 &   1.0 & fh   \\
f214  & J123637.9+621116 & $  13.8^{+   6.2}_{  -5.1}     $ & $   7.5^{+   4.7}_{  -3.5}     $ & $< 10.95                       $ & $< 10.19                       $ & $  0.16^{+  0.04}_{ -0.03}     $ & $  0.02^{+  0.01}_{ -0.01}     $ & $<  0.19                       $ & $<  0.36                       $ &$10^{ -7.2}          $ &   3.10 &  -1.0 & sf   \\
f219  & J123639.2+621656 & $  10.3^{+   6.3}_{  -5.2}     $ & $  11.2^{+   5.6}_{  -4.4}     $ & $< 13.09                       $ & $< 11.36                       $ & $  0.13^{+  0.03}_{ -0.03}     $ & $  0.04^{+  0.01}_{ -0.01}     $ & $<  0.25                       $ & $<  0.43                       $ &$10^{ -5.9}          $ &   3.00 &  -1.0 & s    \\
f230  & J123641.1+622412 & $  79.0^{+  15.8}_{ -14.8}     $ & $  21.0^{+  10.0}_{  -8.9}     $ & $  57.1^{+  13.4}_{ -12.4}     $ & $< 14.15                       $ & $  1.70^{+  0.12}_{ -0.11}     $ & $  0.12^{+  0.02}_{ -0.01}     $ & $  1.90^{+  0.17}_{ -0.15}     $ & $<  0.95                       $ &$10^{ -8.0}          $ &  10.10 &   0.5 & fh   \\
f240  & J123643.2+621626 & $  10.8^{+   5.9}_{  -4.8}     $ & $<  5.33                       $ & $< 11.15                       $ & $< 10.22                       $ & $  0.15^{+  0.04}_{ -0.03}     $ & $<  0.01                       $ & $<  0.23                       $ & $<  0.44                       $ &$10^{ -5.6}          $ &   2.40 &  -1.0 & f    \\
f246  & J123644.4+622556 & $  98.1^{+  20.3}_{ -19.2}     $ & $<  9.46                       $ & $  98.5^{+  17.8}_{ -16.8}     $ & $  45.3^{+  13.4}_{ -12.3}     $ & $  2.20^{+  0.12}_{ -0.11}     $ & $<  0.06                       $ & $  3.40^{+  0.21}_{ -0.20}     $ & $  3.40^{+  0.30}_{ -0.28}     $ &$10^{ -8.0}          $ &  11.80 &   1.0 & fh   \\
f247  & J123644.8+621715 & $<  5.97                       $ & $  12.1^{+   5.6}_{  -4.4}     $ & $< 12.86                       $ & $< 10.66                       $ & $<  0.03                       $ & $  0.05^{+  0.01}_{ -0.01}     $ & $<  0.29                       $ & $<  0.50                       $ &$10^{ -5.4}          $ &   3.10 &  -1.0 & s    \\
f249  & J123645.1+621106 & $  14.7^{+   6.2}_{  -5.1}     $ & $  11.6^{+   5.2}_{  -4.1}     $ & $< 10.63                       $ & $<  9.84                       $ & $  0.19^{+  0.04}_{ -0.04}     $ & $  0.04^{+  0.01}_{ -0.01}     $ & $<  0.20                       $ & $<  0.37                       $ &$10^{ -6.4}          $ &   3.00 &  -1.0 & f    \\
f253  & J123645.3+621522 & $  12.5^{+   5.7}_{  -4.5}     $ & $  13.6^{+   5.3}_{  -4.2}     $ & $< 10.30                       $ & $<  9.47                       $ & $  0.15^{+  0.04}_{ -0.03}     $ & $  0.04^{+  0.01}_{ -0.01}     $ & $<  0.18                       $ & $<  0.34                       $ &$10^{ -5.7}          $ &   1.30 &  -1.0 & f    \\
f255  & J123645.4+621127 & $  10.1^{+   5.6}_{  -4.4}     $ & $  12.7^{+   5.2}_{  -4.1}     $ & $<  9.89                       $ & $<  9.45                       $ & $  0.13^{+  0.04}_{ -0.03}     $ & $  0.04^{+  0.01}_{ -0.01}     $ & $<  0.18                       $ & $<  0.35                       $ &$10^{ -8.0}          $ &   2.70 &  -1.0 & s    \\
f257  & J123645.9+620833 & $  20.8^{+   9.1}_{  -8.0}     $ & $< 15.78                       $ & $  14.6^{+   7.8}_{  -6.8}     $ & $< 16.10                       $ & $  0.26^{+  0.04}_{ -0.03}     $ & $<  0.05                       $ & $  0.28^{+  0.05}_{ -0.04}     $ & $<  0.60                       $ &$10^{ -6.9}          $ &   5.60 &   1.0 & fh   \\
\hline
\end{tabular}
\end{center}
\end{table}
\end{landscape}

\setcounter{table}{1}
\begin{landscape}
\begin{table}
\centering
\caption{Additional sources detected in the HDF-N (continued).
}
\begin{center}
\begin{tabular}{@{}ccrrrrrrrrcccc@{}}
\hline
Cat & CXO HDFN & FB & SB & HB & UB & 0.5-10.0 & 0.5-2.0 & 2-10 & 5-10 & 
  $p_{\rm min}$ & OAA & HR & Flags \\
No. & (2000) & cts & cts & cts & cts & flux & flux & flux & flux & &  ($\prime$) & & \\
(1) & (2) & (3) & (4) & (5) & (6) & (7) & (8) & (9) & (10) & (11) & (12) & (13) & (14)\\
\hline
f263  & J123646.5+621048 & $  10.5^{+   5.9}_{  -4.8}     $ & $   6.2^{+   4.6}_{  -3.4}     $ & $< 11.03                       $ & $< 10.29                       $ & $  0.14^{+  0.04}_{ -0.03}     $ & $  0.02^{+  0.01}_{ -0.01}     $ & $<  0.22                       $ & $<  0.42                       $ &$10^{ -8.0}          $ &   3.30 &  -1.0 & sf   \\
f266  & J123647.5+621639 & $   8.1^{+   5.4}_{  -4.3}     $ & $<  9.19                       $ & $  14.3^{+   5.8}_{  -4.7}     $ & $<  9.11                       $ & $  0.13^{+  0.04}_{ -0.03}     $ & $<  0.04                       $ & $  0.36^{+  0.10}_{ -0.08}     $ & $<  0.45                       $ &$10^{ -6.1}          $ &   2.50 &   1.0 & h    \\
f272  & J123648.3+620246 & $  71.9^{+  13.3}_{ -12.3}     $ & $  49.1^{+  10.0}_{  -9.0}     $ & $<  9.53                       $ & $<  9.53                       $ & $  5.00^{+  0.44}_{ -0.41}     $ & $  0.91^{+  0.11}_{ -0.10}     $ & $<  0.97                       $ & $<  2.10                       $ &$10^{ -8.0}          $ &  11.40 &  -1.0 & fs   \\
f273  & J123648.6+620444 & $  61.0^{+  13.6}_{ -12.6}     $ & $  62.0^{+  11.5}_{ -10.4}     $ & $<  9.53                       $ & $<  9.53                       $ & $  1.80^{+  0.15}_{ -0.14}     $ & $  0.48^{+  0.05}_{ -0.05}     $ & $<  0.40                       $ & $<  0.85                       $ &$10^{ -8.0}          $ &   9.40 &  -1.0 & sf   \\
f282  & J123650.1+621239 & $  14.4^{+   5.8}_{  -4.7}     $ & $<  6.93                       $ & $<  9.11                       $ & $<  8.93                       $ & $  0.17^{+  0.04}_{ -0.04}     $ & $<  0.02                       $ & $<  0.16                       $ & $<  0.31                       $ &$10^{ -5.6}          $ &   1.50 & ... & f    \\
f287  & J123651.2+621551 & $  12.7^{+   5.7}_{  -4.5}     $ & $   9.6^{+   4.8}_{  -3.7}     $ & $< 10.10                       $ & $<  9.37                       $ & $  0.15^{+  0.04}_{ -0.03}     $ & $  0.03^{+  0.01}_{ -0.01}     $ & $<  0.18                       $ & $<  0.34                       $ &$10^{ -5.6}          $ &   1.80 &  -1.0 & s    \\
f302  & J123653.9+621503 & $  12.4^{+   5.6}_{  -4.4}     $ & $   4.2^{+   3.8}_{  -2.6}     $ & $   8.0^{+   4.7}_{  -3.5}     $ & $<  8.32                       $ & $  0.15^{+  0.04}_{ -0.03}     $ & $  0.01^{+  0.01}_{  0.00}     $ & $  0.15^{+  0.05}_{ -0.04}     $ & $<  0.30                       $ &$10^{ -5.8}          $ &   1.20 &   0.3 & f    \\
f303  & J123653.9+621045 & $<  6.07                       $ & $   6.3^{+   4.8}_{  -3.7}     $ & $< 12.82                       $ & $< 11.07                       $ & $<  0.02                       $ & $  0.02^{+  0.01}_{ -0.01}     $ & $<  0.22                       $ & $<  0.39                       $ &$10^{ -5.5}          $ &   3.40 &  -1.0 & s    \\
f321  & J123657.8+620826 & $  20.2^{+   9.8}_{  -8.7}     $ & $  16.2^{+   7.5}_{  -6.4}     $ & $< 12.03                       $ & $< 17.28                       $ & $  0.25^{+  0.03}_{ -0.03}     $ & $  0.05^{+  0.01}_{ -0.01}     $ & $<  0.22                       $ & $<  0.65                       $ &$10^{ -6.1}          $ &   5.80 &  -1.0 & s    \\
f369  & J123707.8+621056 & $  16.9^{+   8.2}_{  -7.1}     $ & $  24.8^{+   7.2}_{  -6.1}     $ & $< 16.02                       $ & $< 13.53                       $ & $  0.20^{+  0.03}_{ -0.03}     $ & $  0.08^{+  0.02}_{ -0.01}     $ & $<  0.28                       $ & $<  0.48                       $ &$10^{ -7.2}          $ &   4.00 &  -1.0 & sf   \\
f370  & J123708.0+621658 & $  22.9^{+   7.9}_{  -6.8}     $ & $  19.5^{+   6.6}_{  -5.5}     $ & $< 14.83                       $ & $< 12.67                       $ & $  0.27^{+  0.05}_{ -0.04}     $ & $  0.06^{+  0.01}_{ -0.01}     $ & $<  0.26                       $ & $<  0.44                       $ &$10^{ -8.0}          $ &   3.70 &  -1.0 & sf   \\
f372  & J123708.3+621513 & $  15.4^{+   6.4}_{  -5.3}     $ & $   6.1^{+   4.6}_{  -3.4}     $ & $   9.7^{+   5.4}_{  -4.3}     $ & $< 10.72                       $ & $  0.19^{+  0.04}_{ -0.04}     $ & $  0.02^{+  0.01}_{ -0.01}     $ & $  0.18^{+  0.05}_{ -0.04}     $ & $<  0.38                       $ &$10^{ -6.1}          $ &   2.70 &   0.2 & f    \\
f390  & J123712.7+621545 & $  19.4^{+   7.2}_{  -6.1}     $ & $  10.2^{+   5.4}_{  -4.3}     $ & $< 13.45                       $ & $< 11.25                       $ & $  0.23^{+  0.04}_{ -0.04}     $ & $  0.03^{+  0.01}_{ -0.01}     $ & $<  0.23                       $ & $<  0.39                       $ &$10^{ -8.0}          $ &   3.40 &  -1.0 & sf   \\
f414  & J123718.7+621314 & $< 15.48                       $ & $   6.3^{+   5.0}_{  -3.8}     $ & $< 13.22                       $ & $< 10.89                       $ & $<  0.25                       $ & $  0.03^{+  0.01}_{ -0.01}     $ & $<  0.31                       $ & $<  0.51                       $ &$10^{ -6.3}          $ &   3.70 &  -1.0 & s    \\
f417  & J123719.8+621336 & $  20.3^{+   7.5}_{  -6.4}     $ & $< 11.86                       $ & $  17.0^{+   6.8}_{  -5.7}     $ & $  12.4^{+   5.9}_{  -4.8}     $ & $  0.28^{+  0.05}_{ -0.04}     $ & $<  0.04                       $ & $  0.37^{+  0.08}_{ -0.06}     $ & $  0.54^{+  0.14}_{ -0.11}     $ &$10^{ -5.6}          $ &   3.80 &   1.0 & u    \\
f444  & J123727.1+621022 & $  26.7^{+  11.3}_{ -10.2}     $ & $< 16.63                       $ & $  23.9^{+  10.0}_{  -8.9}     $ & $  24.4^{+   8.8}_{  -7.7}     $ & $  0.34^{+  0.04}_{ -0.03}     $ & $<  0.06                       $ & $  0.47^{+  0.06}_{ -0.05}     $ & $  1.00^{+  0.15}_{ -0.13}     $ &$10^{ -5.7}          $ &   5.90 &   1.0 & u    \\
f445  & J123727.6+620857 & $  15.6^{+   9.0}_{  -7.9}     $ & $  16.0^{+   7.5}_{  -6.4}     $ & $< 16.63                       $ & $< 16.52                       $ & $  0.44^{+  0.06}_{ -0.05}     $ & $  0.12^{+  0.02}_{ -0.02}     $ & $<  0.69                       $ & $<  1.40                       $ &$10^{ -5.5}          $ &   7.00 &  -1.0 & s    \\
f448  & J123729.9+621300 & $  11.9^{+   8.9}_{  -7.8}     $ & $  17.1^{+   7.5}_{  -6.4}     $ & $< 16.12                       $ & $< 17.09                       $ & $  0.15^{+  0.02}_{ -0.02}     $ & $  0.06^{+  0.01}_{ -0.01}     $ & $<  0.30                       $ & $<  0.64                       $ &$10^{ -5.4}          $ &   5.10 &  -1.0 & s    \\
f454  & J123732.0+620335 & $  15.5^{+   7.4}_{  -6.3}     $ & $  12.9^{+   6.0}_{  -4.9}     $ & $< 14.56                       $ & $< 11.58                       $ & $  2.70^{+  0.49}_{ -0.42}     $ & $  0.58^{+  0.14}_{ -0.12}     $ & $<  3.70                       $ & $<  6.40                       $ &$10^{ -6.2}          $ &  11.80 &  -1.0 & s    \\
f455  & J123732.2+621303 & $  35.2^{+  10.6}_{  -9.6}     $ & $  30.7^{+   8.6}_{  -7.5}     $ & $< 11.58                       $ & $< 17.32                       $ & $  0.44^{+  0.05}_{ -0.05}     $ & $  0.10^{+  0.02}_{ -0.01}     $ & $<  0.21                       $ & $<  0.64                       $ &$10^{ -6.7}          $ &   5.30 &  -1.0 & s    \\
f459  & J123734.4+622326 & $  52.4^{+  14.9}_{ -13.8}     $ & $  39.4^{+  10.8}_{  -9.7}     $ & $  18.9^{+  11.6}_{ -10.6}     $ & $< 13.05                       $ & $  2.20^{+  0.17}_{ -0.16}     $ & $  0.44^{+  0.05}_{ -0.04}     $ & $  1.30^{+  0.13}_{ -0.12}     $ & $<  1.90                       $ &$10^{ -6.3}          $ &  10.80 &  -0.3 & f    \\
f464  & J123735.4+621055 & $  54.6^{+  12.9}_{ -11.9}     $ & $  35.0^{+   9.8}_{  -8.8}     $ & $< 18.68                       $ & $< 18.68                       $ & $  0.79^{+  0.07}_{ -0.07}     $ & $  0.13^{+  0.02}_{ -0.02}     $ & $<  0.40                       $ & $<  0.82                       $ &$10^{ -6.1}          $ &   6.40 &  -1.0 & f    \\
f474  & J123738.5+621702 & $< 19.28                       $ & $  22.7^{+   9.1}_{  -8.0}     $ & $< 19.28                       $ & $< 19.28                       $ & $<  0.26                       $ & $  0.08^{+  0.01}_{ -0.01}     $ & $<  0.38                       $ & $<  0.78                       $ &$10^{ -5.9}          $ &   6.60 &  -1.0 & s    \\
f479  & J123740.8+621231 & $  38.0^{+  12.6}_{ -11.5}     $ & $  25.5^{+   9.6}_{  -8.5}     $ & $< 19.28                       $ & $< 19.28                       $ & $  0.48^{+  0.04}_{ -0.04}     $ & $  0.09^{+  0.01}_{ -0.01}     $ & $<  0.36                       $ & $<  0.73                       $ &$10^{ -6.3}          $ &   6.40 &  -1.0 & s    \\
f483  & J123741.8+620617 & $  42.4^{+  11.7}_{ -10.7}     $ & $< 18.66                       $ & $  31.4^{+  10.0}_{  -9.0}     $ & $< 18.63                       $ & $  1.70^{+  0.17}_{ -0.16}     $ & $<  0.19                       $ & $  2.00^{+  0.24}_{ -0.22}     $ & $<  2.40                       $ &$10^{ -5.6}          $ &  10.10 &   1.0 & f    \\
f490  & J123745.3+620927 & $  21.1^{+   8.5}_{  -7.5}     $ & $< 14.45                       $ & $< 17.26                       $ & $< 13.94                       $ & $  0.72^{+  0.11}_{ -0.10}     $ & $<  0.13                       $ & $<  0.88                       $ & $<  1.50                       $ &$10^{ -5.5}          $ &   8.20 & ... & f    \\
f501  & J123755.6+621507 & $ 164.4^{+  19.4}_{ -18.4}     $ & $ 158.5^{+  16.9}_{ -15.9}     $ & $  32.0^{+  12.9}_{ -11.9}     $ & $< 18.65                       $ & $  2.20^{+  0.13}_{ -0.12}     $ & $  0.57^{+  0.04}_{ -0.04}     $ & $  0.66^{+  0.06}_{ -0.05}     $ & $<  0.76                       $ &$10^{ -8.0}          $ &   8.00 &  -0.6 & fs   \\
f504  & J123758.6+621643 & $  45.7^{+  18.5}_{ -17.4}     $ & $  15.1^{+  13.4}_{ -12.3}     $ & $< 14.52                       $ & $< 18.65                       $ & $  0.62^{+  0.04}_{ -0.04}     $ & $  0.05^{+  0.00}_{  0.00}     $ & $<  0.03                       $ & $<  0.77                       $ &$10^{ -5.5}          $ &   8.70 &   0.2 & f    \\
f509  & J123801.0+621111 & $  49.2^{+  11.7}_{ -10.7}     $ & $  40.6^{+   9.8}_{  -8.8}     $ & $< 18.65                       $ & $< 17.31                       $ & $  1.70^{+  0.17}_{ -0.16}     $ & $  0.37^{+  0.05}_{ -0.04}     $ & $<  0.95                       $ & $<  1.90                       $ &$10^{ -8.0}          $ &   9.00 &  -1.0 & fs   \\
f512  & J123804.0+621006 & $  36.2^{+  11.4}_{ -10.3}     $ & $< 17.31                       $ & $  33.8^{+  10.0}_{  -9.0}     $ & $< 18.02                       $ & $  1.30^{+  0.14}_{ -0.13}     $ & $<  0.17                       $ & $  1.90^{+  0.24}_{ -0.21}     $ & $<  2.10                       $ &$10^{ -6.0}          $ &   9.80 &   1.0 & h    \\
f519  & J123816.2+620904 & $  44.5^{+  12.4}_{ -11.3}     $ & $  21.5^{+   9.1}_{  -8.0}     $ & $  22.2^{+   9.6}_{  -8.6}     $ & $< 17.42                       $ & $  2.40^{+  0.23}_{ -0.21}     $ & $  0.31^{+  0.04}_{ -0.04}     $ & $  1.90^{+  0.24}_{ -0.22}     $ & $<  3.10                       $ &$10^{ -8.0}          $ &  11.50 &   0.1 & f    \\
f522  & J123817.8+620857 & $  54.8^{+  12.1}_{ -11.1}     $ & $  32.3^{+   9.4}_{  -8.3}     $ & $  22.9^{+   9.1}_{  -8.0}     $ & $< 17.06                       $ & $  3.90^{+  0.38}_{ -0.35}     $ & $  0.60^{+  0.08}_{ -0.07}     $ & $  2.50^{+  0.36}_{ -0.32}     $ & $<  4.00                       $ &$10^{ -8.0}          $ &  11.70 &  -0.1 & fs   \\
f524  & J123819.9+620934 & $  46.5^{+  12.5}_{ -11.5}     $ & $  27.6^{+   9.5}_{  -8.5}     $ & $< 17.06                       $ & $< 17.06                       $ & $  2.50^{+  0.23}_{ -0.21}     $ & $  0.39^{+  0.05}_{ -0.05}     $ & $<  1.30                       $ & $<  3.00                       $ &$10^{ -5.7}          $ &  11.70 &  -1.0 & fs   \\
f529  & J123826.9+621545 & $  66.7^{+  16.6}_{ -15.6}     $ & $< 17.81                       $ & $  57.6^{+  14.3}_{ -13.3}     $ & $< 17.81                       $ & $  2.60^{+  0.18}_{ -0.17}     $ & $<  0.19                       $ & $  3.50^{+  0.28}_{ -0.26}     $ & $<  2.20                       $ &$10^{ -5.6}          $ &  11.70 &   1.0 & f    \\
f531  & J123827.8+621631 & $  70.2^{+  17.0}_{ -15.9}     $ & $  29.7^{+  10.8}_{  -9.7}     $ & $  43.0^{+  14.1}_{ -13.0}     $ & $<  8.81                       $ & $  2.80^{+  0.19}_{ -0.18}     $ & $  0.32^{+  0.04}_{ -0.03}     $ & $  2.70^{+  0.22}_{ -0.21}     $ & $<  1.20                       $ &$10^{ -7.2}          $ &  11.90 &   0.2 & f    \\
f532  & J123830.1+621019 & $  28.3^{+  11.0}_{  -9.9}     $ & $  30.2^{+   9.1}_{  -8.0}     $ & $<  8.81                       $ & $< 17.44                       $ & $  2.20^{+  0.24}_{ -0.22}     $ & $  0.62^{+  0.09}_{ -0.08}     $ & $<  1.00                       $ & $<  4.40                       $ &$10^{ -8.0}          $ &  12.60 &  -1.0 & sf   \\
f536  & J123842.5+621143 & $  69.6^{+  13.5}_{ -12.5}     $ & $  57.0^{+  10.9}_{  -9.8}     $ & $< 16.53                       $ & $< 19.05                       $ & $  5.30^{+  0.46}_{ -0.42}     $ & $  1.20^{+  0.13}_{ -0.12}     $ & $<  1.90                       $ & $<  4.80                       $ &$10^{ -8.0}          $ &  13.60 &  -1.0 & fs   \\
\hline
\end{tabular}
\end{center}
\end{table}
\end{landscape}

\end{document}

\begin{sidewaystable*}
\centering
\begin {tabular}{ccccccccccr}
\multicolumn{11}{l}{{\bf Table 1.} Chandra sources in the ELAIS N1 field.} \\
\hline
&  & {\bf RA} & {\bf Dec} & {\bf Err} & {\bf Net} &  & 
\multicolumn{3}{c}{\bf Flux ($\times 10^{-14}$ erg cm$^{-2}$ s$^{-1}$)}  
& \\
{\bf ID} & {\bf CXOEN1} & {\bf (J2000)} & {\bf (J2000)} & 
{\bf (arcsec)} & {\bf Cts} & {\bf S/N} & {\bf (0.5--8keV)}  
& {\bf (0.5--2keV)} & {\bf (2--8keV)} & {\bf HR\hspace{0.6cm}} \\
\hline
\hline

\tiny
\centering
\caption{Chandra GWS X-ray catalogue (continued)}
\begin{center}
\begin{tabular}{ccrrrrrrrrcccc}
\hline
Cat & CXO GWS & FB & SB & HB & UB & 0.5-10 & 0.5-2.0 & 2-10 & 5-10 & 
  $p_{\rm min}$ & HR & OAA & Flags \\
No. & (2000) & cts & cts & cts & cts & flux & flux & flux & flux & & & (arcmin) \\
(1) & (2) & (3) & (4) & (5) & (6) & (7) & (8) & (9) & (10) & (11) & (12) \\
\hline
\input{psc_raw2.tex}
\hline
\end{tabular}
\end{center}
\end{table*}

\setcounter{table}{1}
\begin{table*}
\tiny
\centering
\caption{Chandra GWS X-ray catalogue (continued)}
\begin{center}
\begin{tabular}{ccrrrrrrrrcccc}
\hline
Cat & CXO GWS & FB & SB & HB & UB & 0.5-10 & 0.5-2.0 & 2-10 & 5-10 & 
  $p_{\rm min}$ & HR & OAA & Flags \\
No. & (2000) & cts & cts & cts & cts & flux & flux & flux & flux & & & ($\prime$) \\
(1) & (2) & (3) & (4) & (5) & (6) & (7) & (8) & (9) & (10) & (11) & (12) \\
\hline
\input{psc_raw3.tex}
\hline
\end{tabular}
\end{center}
\end{table*}

\end{document}